\documentclass{WileyMSP-template}
\usepackage{graphicx}  
\usepackage{dcolumn}   
\usepackage{bm}        
\usepackage{verbatim}   
\usepackage{epstopdf}
\usepackage{footnote,cite}
\usepackage{epsfig}
\usepackage{subfigure}
\usepackage{amssymb}
\usepackage{amsmath,bm}
\usepackage{xcolor}
\usepackage{float}
\usepackage{subfigure}
\usepackage[most]{tcolorbox}
\usepackage{soul}
\usepackage{color}

\begin{document}


\title{Lightweight Low-Noise Linear Isolator Integrating Phase-Engineered Temporal Loops}

\maketitle


\author{Sajjad Taravati}
\author{George V. Eleftheriades}



\begin{affiliations}
The Edward S. Rogers Sr. Department of Electrical and Computer Engineering, University of Toronto, Toronto, Ontario M5S 3G4, Canada\\
Email Address: sajjad.taravati@utoronto.ca
\end{affiliations}



\begin{abstract}
The quest for efficient and versatile microwave and optical isolators has recently spawned several novel space-time-modulated isolator structures. However, such space-time isolators suffer from a large profile and complex architecture caused by progressive nonreciprocal space-time coupling properties. To overcome these limitations, we propose a nonmagnetic phase-engineered temporal-loop-based isolator featuring large isolation levels, weak undesired time harmonics, and a low profile. The proposed isolator is composed of two temporal loops that provide desired constructive and destructive interferences of different time harmonics. Furthermore, these two loops are designed in a way to assure that the circulation and reflection of different time harmonics strengthen a low insertion loss unidirectional signal transmission. An experimental demonstration of the proposed time-modulated isolator at microwave frequencies is provided, featuring strong unidirectional wave transmission through the isolator with more than 27 dB contrast between the forward and backward waves across a fractional bandwidth of $14.3\%$. The proposed isolator outperforms the nonlinear-based and transistor-based isolators by featuring a highly linear response with OP$_\text{1dB}$ of higher than $31$~dBm, high power rating of more than $47$~dBm, and a low noise figure of $3.4$~dB.
\end{abstract}


\section{Introduction}

Isolators are components that block electromagnetic wave signals in one direction but allow them to pass in the opposite direction. Such a critical task is required in numerous applications, including protecting microwave, millimeter wave, terahertz and optical sources from back reflections disturbing the source operation, or mitigating multi-path interference in communication systems. Additionally, isolators generally improve the designability of the overall system as they eliminate spurious interferences, interactions between different components and undesired signal routing. Isolators can be realized by the simultaneous breaking of time-reversal symmetry
and spatial inversion~\cite{SOOHOO_TM_1968,nagulu2018nonreciprocal,Taravati_Kishk_MicMag_2019}. Conventional techniques and structures for the realization of isolators include magnetically biased two-dimensional electron
gas systems~\cite{Hirota_TED_1971}, gyroelectric waveguides~\cite{jawad2017millimeter}, transistor-loaded transmission lines~\cite{lee20055,chang2015design,Taravati_2017_NR_Nongyro,wang2019highly,taravati2021programmable,wang2020210,taravati2021full}, magnetic ferrites~\cite{SOOHOO_TM_1968,tsutsumi1987dielectric,seewald2010ferrite,wu2012novel,cheng2014narrowband,farooqui2014inkjet,marynowski2018integrated,ghaffar2019theory}, nonlinearity~\cite{sounas2018broadband}, and space-time-modulation~\cite{Fan_PRL_109_2012,Wang_TMTT_2014,Taravati_PRB_2017,Taravati_PRB_SB_2017}. Although these approaches have their own unique advantages and applications, they suffer from distinct limitations and disadvantages that restrict their applications. For instance, magnetically biased isolators require bulky magnets and are incompatible with
integrated circuit technology~\cite{Lax_1962}. An alternative approach is transistor-based isolators~\cite{Taravati_2017_NR_Nongyro,taravati2021programmable,taravati2021full} which eliminate the bulky magnet and are compatible with integrated-circuit technology but suffer from limited power handling and noise performance, and are of limited availability at high frequencies. Nonlinear isolators may be a good choice for some applications but they only provide isolation to high power signals, while they pass low-level signals quasi-reciprocally~\cite{shi2015limitationsNL}.

Time modulation has recently attracted a surge of interest thanks to its distinctive capabilities in non-magnetic nonreciprocal signal transmission and frequency generation~\cite{fang2012photonic,estep2014magnetic,Taravati_Kishk_MicMag_2019,taravati_PRApp_2019,ramaccia2021temporal,elnaggar2020modeling,Engheta_PRB_2021,taravati2021pure}. As a result, diversified applications of time modulation techniques have recently been discovered, including signal amplification~\cite{tien1958traveling,tien1958parametric,zhu2020tunable}, unidirectional beam splitting~\cite{Taravati_Kishk_PRB_2018}, pure frequency conversion~\cite{Taravati_PRB_Mixer_2018}, multi-functional and nonreciprocal metasurfaces~\cite{zang2019nonreciprocal_metas,taravati2020full,salary2020time,taravati2020four,wang2020nonreciprocity}, circulators~\cite{estep2014magnetic,nagulu2020multi}, nonreciprocal filters~\cite{wu2020frequency}, and multi-functional antennas~\cite{zang2019nonreciprocal,Taravati_LWA_2017,Taravati_AMA_PRApp_2020}. The linear non-magnetic nonreciprocity undergoing time modulation has been shown to be a prominent alternative to conventional cumbersome ferrite-based, transistor-loaded and nonlinear nonreciprocal devices~\cite{bhandare2005novel,fang2013experimental,Wang_TMTT_2015,Taravati_PRB_SB_2017,correas2019plasmonic,Taravati_PRAp_2018,li2019nonreciprocal,zang2019nonreciprocal}. Such dynamic meta-structures may be designed to introduce linear, highly efficient and reconfigurable isolators.

The advent and development of space-time-modulated isolators over the past decade has opened up an intriguing research domain for the realization of versatile and efficient electromagnetic isolation at different frequencies. As a result of such discoveries, several unique space-time-modulated isolators have been proposed ranging from acoustics~\cite{li2019nonreciprocal,zhu2020tunable} and microwaves~\cite{Wang_TMTT_2015,Taravati_PRB_2017,Taravati_PRB_SB_2017} to optics and photonics~\cite{Fan_PRL_109_2012}. However, the isolation provided by the space-time modulation requires an optically long structure so that an efficient progressive unidirectional coupling between the fundamental space-time harmonic and higher-order space-time harmonics occurs~\cite{Fan_PRL_109_2012,Taravati_PRB_2017,Taravati_PRB_SB_2017}. This in turn leads to a complex and cumbersome architecture, which results in a complex and costly fabrication. From the practical point of view, a space-time-modulated isolator should be implemented using a set of subwavelength-spaced varactors emulating a homogeneous space-time medium, where the length of the medium is in the order of the temporal modulation frequency so that a desired unidirectional coherency between the incident wave signal and the modulation signal occurs. As a consequence, a space-time-modulated isolator may require more than forty varactors~\cite{Wang_TMTT_2015,Taravati_LWA_2017,Taravati_PRB_2017,Taravati_PRB_SB_2017}, which leads to long and cumbersome devices.

Here, we propose a lightweight, low-profile, and non-magnetic isolator formed by two time-modulated loops, each of which comprising phase-shifted temporal transmission lines. The lightweight and low-profile properties originate from the fact that it requires only four varactors. This is in contrast to space-time-modulated isolators where a progressive unidirectional coherence is required; Here the isolation is achieved by proper constructive and destructive interferences at the two ports of the isolator. We show that by proper design of the band structure of these phase-shifted temporal transmission lines, a proper signal isolation between the forward and backward waves can be achieved. The frequency and phase transition in a time-varying transmission line is achieved by varactors excited by a time-varying voltage. The isolator possesses a lightweight and low-profile structure, and exhibits more than 32 dB isolation with very weak spurious time harmonics. Such promising results show that this isolator can be used as a suitable alternative to conventional bulky magnetic microwave isolators.

\section{Results and Discussion}

\begin{figure}
	\centering
	\includegraphics[width=130mm]{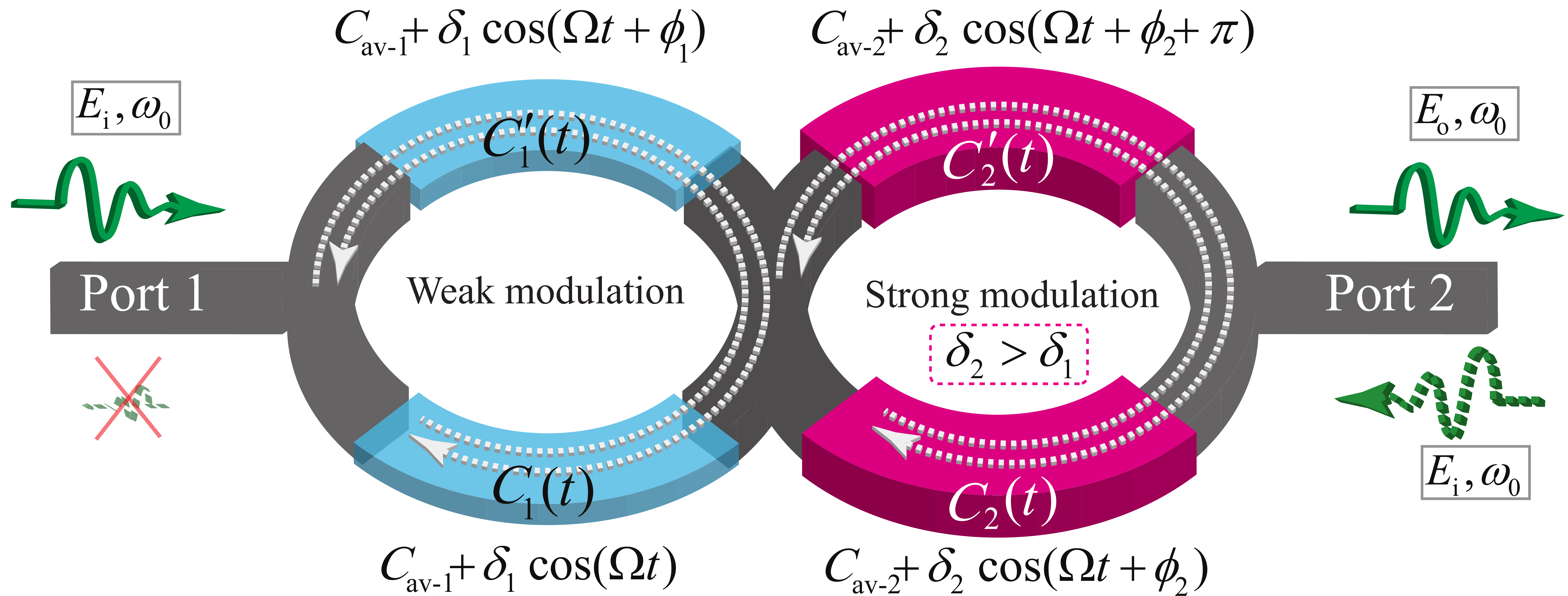}
	\caption{Schematic representation of the magnetless isolator formed by two time-modulated loops.}
	\label{fig:sch}
\end{figure}

Figure~\ref{fig:sch} depicts the architecture of the time-modulated isolator. This isolator is formed by two time-varying loops, themselves composed of two phase-shifted temporal transmission lines. The left temporal loop is composed of two (upper and lower) temporal transmission lines, where the capacitance of the transmission lines is modulated in time, incurring a $\phi_1$ phase difference between the upper and lower temporal transmission lines, i.e.,

\begin{subequations}
	\begin{equation}\label{eqa:C1}
	C_1(t)=C_\text{av-1}+\delta_1 \cos(\Omega t),
	\end{equation}
	\begin{equation}\label{eqa:C1p}
	C_1'(t)=C_\text{av-1}+\delta_1 \cos(\Omega t+\phi_1).
	\end{equation}

	Furthermore, the right temporal loop is composed of two (upper and lower) temporal transmission lines, where the capacitance of the transmission lines is modulated in time and incurring a $\pi$ phase difference between the upper and lower temporal transmission lines, i.e., 	
	\begin{equation}\label{eqa:C2}
	C_2(t)=C_\text{av-2}+\delta_2 \cos(\Omega t+\phi_2),
	\end{equation}
	\begin{equation}\label{eqa:C2p}
	C_2'(t)=C_\text{av-2}+\delta_2 \cos(\Omega t+\phi_2+\pi).
	\end{equation}
\end{subequations}

In~\eqref{eqa:C1} to~\eqref{eqa:C2p}, $C_\text{av-1}$ and $C_\text{av-2}$ are the average permitivitties of the first and second loops, respectively, $\delta_{1}$ and $\delta_{2}$ denote the modulation amplitudes of the first and second loops, respectively, and $\Omega$ is the temporal modulation frequency. Here, the right temporal loop assumes a stronger modulation amplitude, that is, $\delta_2>>\delta_1$. We will show that by proper design of the phase and amplitude of the time modulation of the two loops, an appropriate constructive interference for the forward wave incidence (left to right propagation) and a proper destructive interference for the backward wave incidence (right to left propagation) may be achieved through the three (gray-colored in Fig.~\ref{fig:sch}) nodes of the architecture, i.e., at the input, middle and output of the structure.

\begin{figure}
	\centering
	\includegraphics[width=0.9\columnwidth]{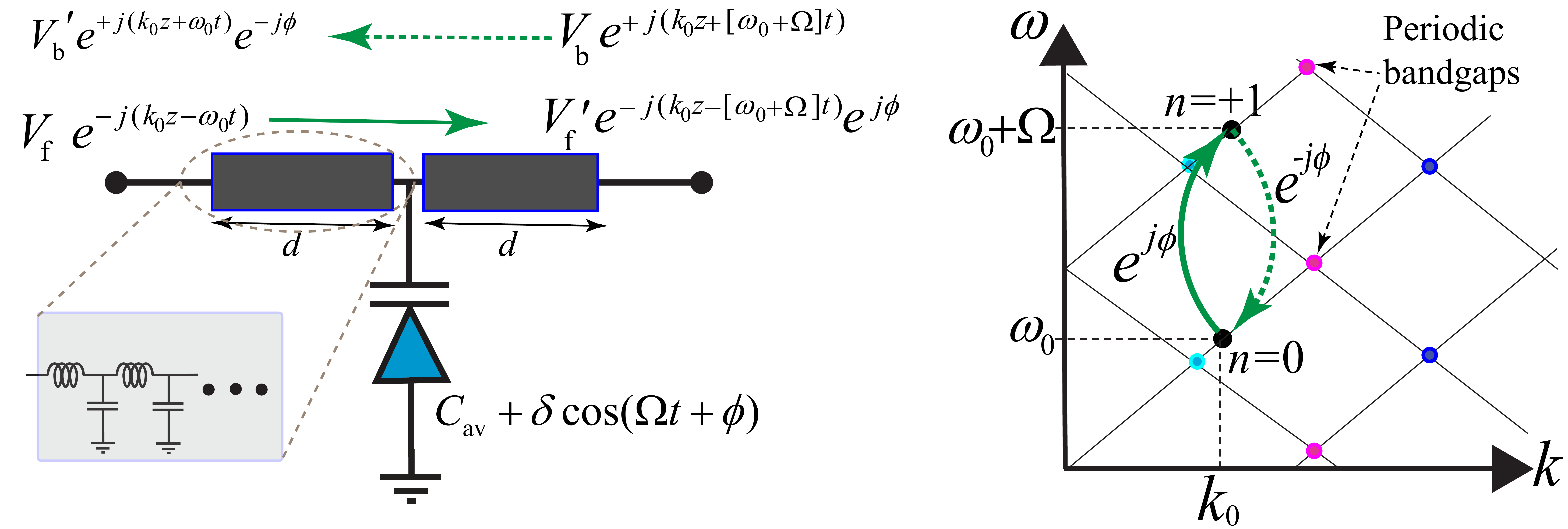}
	\caption{Forward and backward frequency-phase transitions in a temporal transmission line.}
	\label{fig:disp1}
\end{figure}

To best understand the effect of the phase and amplitude of the time modulation on the output voltage of the temporal transmission lines, we first investigate the variation of the voltage through a temporal-varactor-loaded transmission line. Figure~\ref{fig:disp1} illustrates a transmission line loaded by a phase-shifted temporal varactor. The time-varying capacitance of the transmission line is expressed as
\begin{equation}\label{eqa:permit}
C_\text{eq}(t)=C_\text{av} + \delta \cos(\Omega t+\phi),
\end{equation}
where $C_\text{av}$ is the average capacitance of the time-varying varactor. The total capacitance of the transmission line reads $C_\text{tot}(t)=C_0C_\text{eq}(t)$, where $C_0$ is the mean capacitance per unit length of the transmission line. It may be shown that, by proper engineering of the dispersion of the transmission line, the output voltage acquires a frequency transition (up/down-conversion) accompanied by a phase transition. Such frequency and phase transitions are shown in the dispersion diagram in the right side of Fig.~\ref{fig:disp1}, where a frequency up-conversion (from $\omega_0$ to $\omega_0+\Omega$) is accompanied by the phase addition of $\phi$.

We assume the dispersion bands of the structure are engineered so that only the fundamental and the first higher-order time-harmonics are excited, whereas all higher-order time-harmonics are suppressed. Hence, the voltage is defined based on the superposition of the $n=0$ and $n=-1$ space-time harmonics fields, i.e.,
\begin{equation}\label{eqa:el}
V_\text{S}(z,t)=a_{0}(z) e^{-i \left(k_0 z -\omega_0 t \right)}+a_{1}(z) e^{-i \left(k_0 z -(\omega_0+\Omega) t \right)}.
\end{equation}
Since the transmission line is modulated only in time, and not in space, we consider an identical wave number $k_0$ for both the fundamental and the higher-order harmonics. As it is shown in Fig.~\ref{fig:disp1}, and as a result of the temporal sinusoidally periodic capacitance (in Eq.~\eqref{eqa:permit}) the dispersion diagram of the temporal transmission line is periodic with respect to $\omega$ axis~\cite{Halevi_PRA_2009}, that is $k(\omega_0+\Omega)=k(\omega_0)=k_0$. As a result of such temporal periodicity, electromagnetic transitions are vertical in the dispersion diagram (Fig.~\ref{fig:disp1}). The unknown spatially variant amplitudes $a_{0}(z)$ and $a_{1}(z)$ are to be found through satisfying both the Telegrapher’s equations and initial conditions at $z=0$. Considering a lossless transmission line with time-varying capacitance, the Telegrapher’s equations read%
\begin{subequations}
	\begin{equation}\label{eqa:T1}
	\frac{\partial V(z,t) }{\partial z}=-L_0 \frac{\partial I(z,t) }{\partial t},
	\end{equation}
	\begin{equation}\label{eqa:T2}
	\frac{\partial I(z,t) }{\partial z}=-C_0 \frac{\partial [ C_\text{eq}(t) V(z,t)] }{\partial t},
	\end{equation}
\end{subequations}
where $L_0$ and $C_0$ are the inductance and capacitance per unit length of the unloaded transmission line,
respectively. Equations~\eqref{eqa:T1} and~\eqref{eqa:T2} yield
\begin{align}\label{eqa:wave_eq}
\frac{\partial^{2} V(z,t)}{\partial z^{2}}= \dfrac{1}{v_\text{p}^2} \frac{\partial^{2} [C_\text{eq}(t) V(z,t)]}{\partial t^{2}},
\end{align}
where $v_\text{p}=1/\sqrt{L_0 C_0}$.

The up-conversion assumes the initial conditions of $a_{0}(0)=V_0$ and $a_{1}(0)=0$, which gives (see Supplementary Information)
\begin{subequations}
	\begin{equation}\label{eq:up1}
	a_{0}(z)=V_0 \cos\left(\dfrac{\delta k_1}{4 C_\text{av}} z\right),
	\end{equation}
	\begin{equation}\label{eq:up2}
	a_{1}(z)=V_0\dfrac{k_1 }{k_0} e^{+i \phi}  \sin\left(\dfrac{\delta k_1}{4 C_\text{av}} z\right).
	\end{equation}
\end{subequations}

Equations~\eqref{eq:up1} and~\eqref{eq:up2} show that the change of the frequency and phase of the up-converted signal corresponds to the frequency and phase of the temporal modulation signal, $\Omega$ and $+\phi$. In addition, the maximum amplitude of the up-converted signal $a_1(z)$ occurs at $\delta k_1/ C_\text{av}=2\pi$. This reveals that one can control the amplitude of the up-converted signal by changing the modulation amplitude $\delta$. Following the same procedure, we can determine the input and output voltages for the down-conversion. The down-conversion assumes the initial conditions of $a_{0}(0)=0$ and $a_{1}(0)=V_1=\text{Max}[a_{1}(z)]=V_0k_1 /k_0 exp(+i \phi_0)$, which gives
\begin{subequations}
	\begin{equation}\label{eq:down1}
	a_{1}(z)=V_1 \cos\left(\dfrac{\delta k_1}{4 C_\text{av}} z\right),
	\end{equation}
	and
	\begin{equation}\label{eq:down2}
	a_{0}(z)= V_1\dfrac{k_0 }{k_1} e^{-i \phi}  \sin\left(\dfrac{\delta k_1}{4 C_\text{av}} z \right)=V_0  e^{i (\phi_0-\phi)}  \sin\left(\dfrac{\delta k_1}{4 C_\text{av}} z \right).
	\end{equation}
\end{subequations}

Equations~\eqref{eq:down1} and~\eqref{eq:down2} show that the change of the frequency and phase of the down-converted signal correspond to the frequency and phase of the temporal modulation signal, $\Omega$ and $-\phi$.

\begin{figure}
	\centering
	\includegraphics[width=160mm]{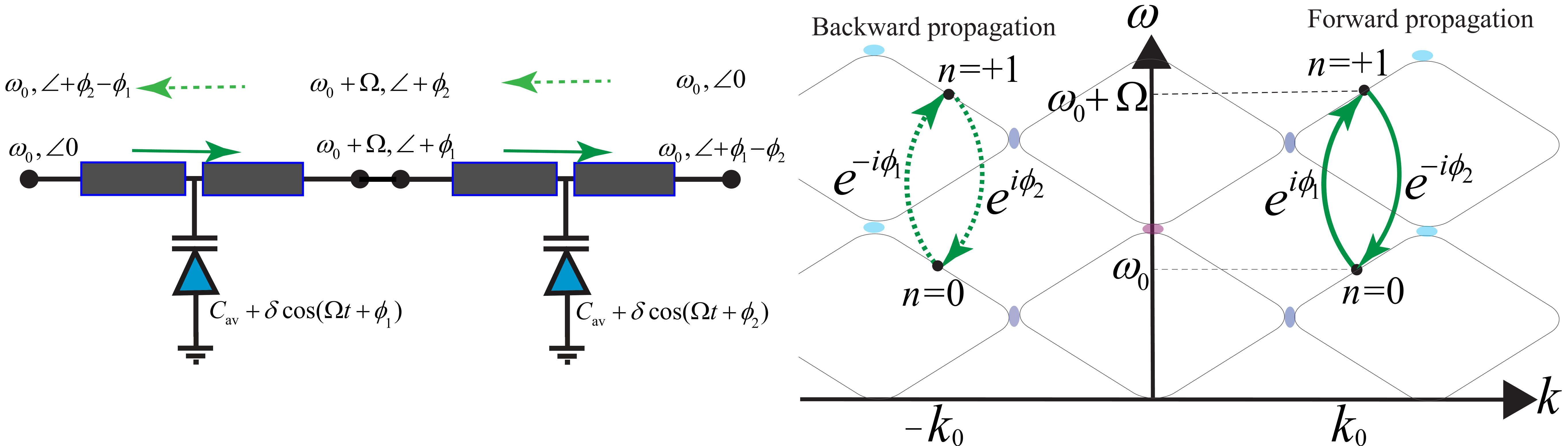}
	\caption{Schematic representation of forward and backward frequency-phase transitions in two cascaded temporal transmission line sections.}
	\label{fig:disp2}
\end{figure}

Figure~\ref{fig:disp2} shows a schematic representation of the dispersion diagram for asymmetric frequency-phase transitions in a temporal transmission line. Here, a frequency up-conversion from $\omega_0$ to $\omega_0+\Omega$ is accompanied by a positive additive phase, i.e., $+\phi_1$ in the forward transmission and $+\phi_2$ in the backward transmission. However, a frequency down-conversion from $\omega_0+\Omega$ to $\omega_0$ is accompanied by a negative additive phase, i.e., $-\phi_2$ in the forward transmission and $-\phi_1$ in the backward transmission.

\begin{figure}
	\begin{center}
		\subfigure[]{\label{fig:p1} 
			\includegraphics[width=0.7\columnwidth]{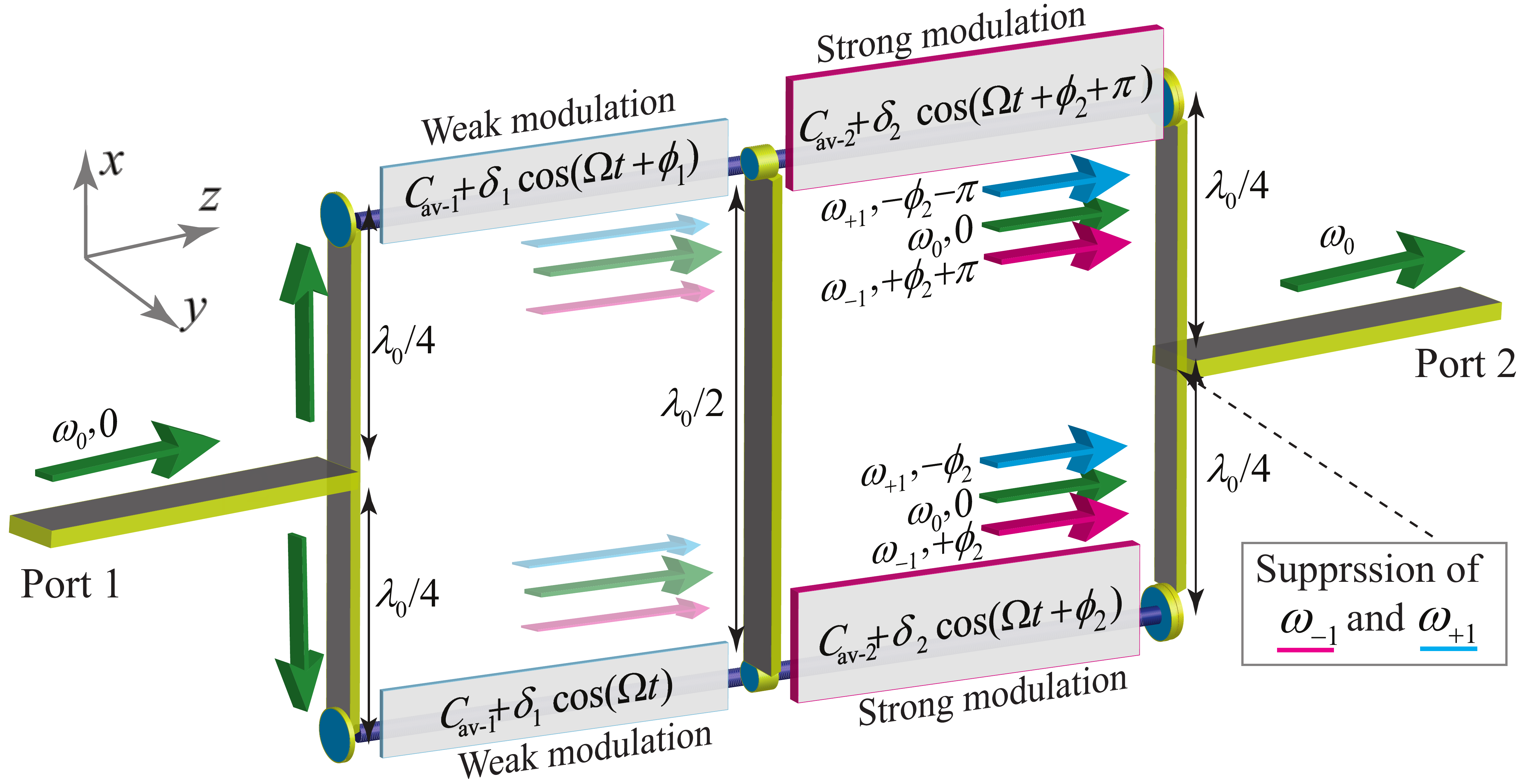}} 
		\subfigure[]{\label{fig:p2} 
			\includegraphics[width=0.7\columnwidth]{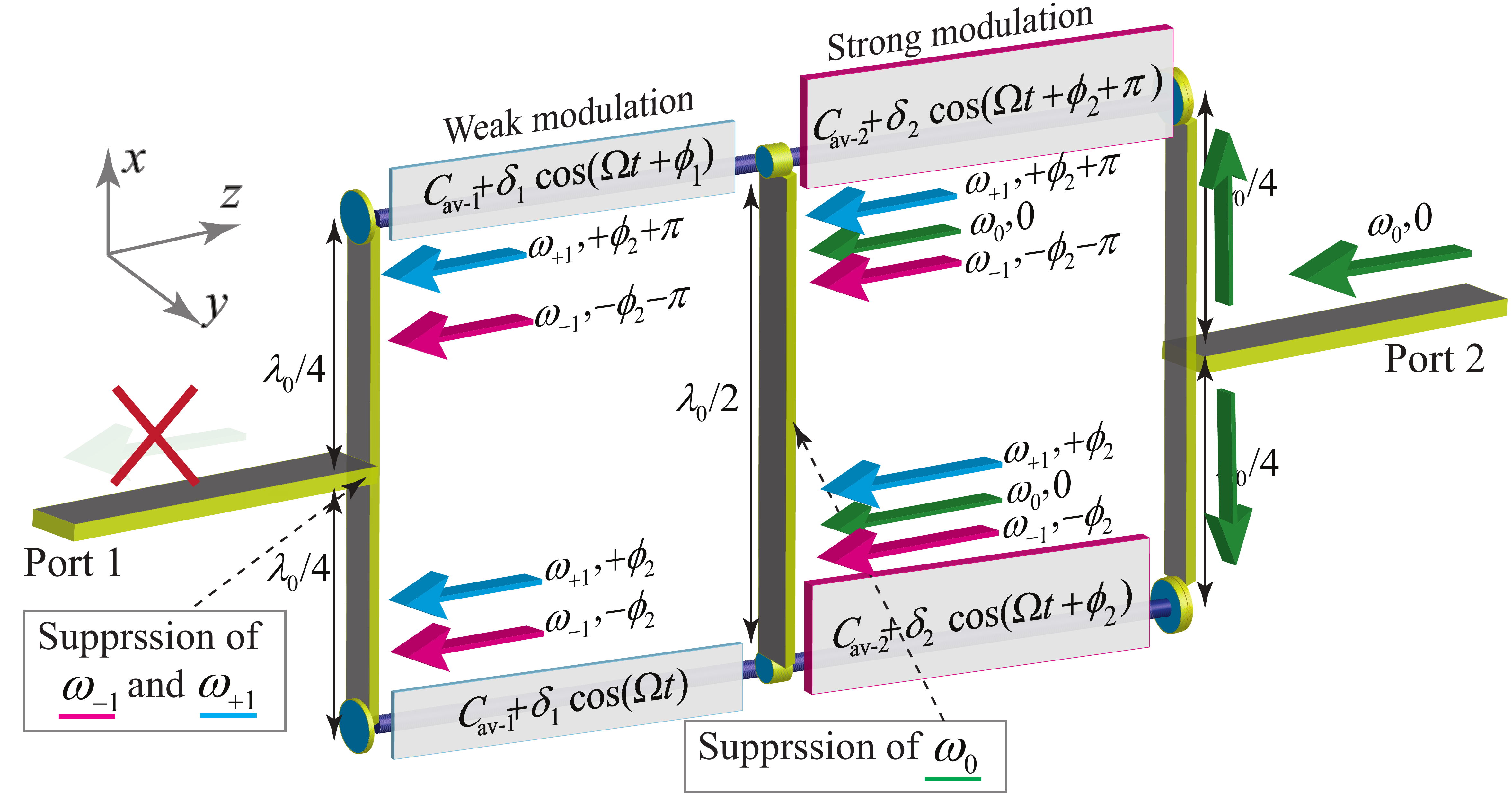}}
		\caption{Engineered frequency and phase transitions in the proposed temporal isolator leading to unidirectional signal transmission. (a) Forward signal injection, where the injected signal passes through the isolator. (b) Backward signal injection, where the injected signal cannot pass through the isolator.} 
		\label{Fig:pp}
	\end{center}
\end{figure}

Figures~\ref{fig:p1} sketches a general representation of the architecture of the proposed time-varying isolator. The isolator is composed of two temporal loops, each of which formed by two sinusoidally time-varying transmission lines. In order to achieve the desired constructive and destructive interferences at the two ports of the isolator, for both forward and backward signal injections, two power splitters/combiners possessing two quarter-wavelength arms are utilized at the two ends of the isolator. In addition, a half-wavelength transmission line is placed at the middle of the isolator to separate the two loops from each other and provide desired constructive and destructive interferences at different time harmonics. We aim to achieve a unidirectional signal transmission by proper engineering of the forward (Figures~\ref{fig:p1}) and backward (~\ref{fig:p2}) frequency and phase transitions in the proposed temporal isolator. In the forward signal transmission in Fig.~\ref{fig:p1}, the signal at $\omega_0$ is injected to port 1 of the isolator, passes through the two loops of the isolator and generates first higher-order time harmonics. Here, the fundamental time harmonic $\omega_0$ at the upper and lower arms of the right-side loop acquire identical phases, and therefore combine coherently at port 2 of the isolator. In contrast, the first higher-order time harmonics acquire different phase shifts at the upper and lower arms of the right-side loop, i.e., with $\pi$ phase difference, and hence produce a null at port 2 of the isolator. As a result, due to the constructive interference of the fundamental time harmonic $\omega_0$, and destructive interference of the first higher-order time harmonics ($\omega_{+1}$ and $\omega_{-1}$) at the common port of the right-side power combiner, a full transmission of the fundamental time harmonic occurs while undesired harmonics are highly suppressed.

For the backward signal injection, shown in Fig.~\ref{fig:p2}, the input signal at $\omega_0$ is injected to port 2 of the isolator (on the right side of the isolator). The injected signal passes through the right side loop of the isolator and generates first higher-order time harmonics but with different phases at the upper and lower arms of the isolator, i.e., with $\pi$ phase difference for the first higher-order time harmonics ($\omega_{+1}$ and $\omega_{-1}$), while the fundamental time harmonic $\omega_0$ at the upper and lower arms of the right-side loop acquire identical phases. As a consequence, a destructive interference and suppression of the fundamental time harmonic $\omega_0$ occurs by the half-wavelength long middle interconnector. One should note that since the $\omega_{+1}$ and $\omega_{-1}$ time harmonics acquire $\pi$ phase difference at the upper and lower arms of the right-side loop, the half-wavelength long middle interconnector provides a constructive interference for them. However, these two harmonics ($\omega_{+1}$ and $\omega_{-1}$) will generate a complete null at the port 1 of the isolator (output port for backward transmission). We shall stress that, as a result of the large asymmetry in the amplitudes of the two loops of the isolator ($\delta_1<<\delta_2$), the $\omega_0$ harmonic that is suppressed by the half-wavelength long middle interconnector in the backward transmission, cannot be regenerated by the left-side loop which possesses a small modulation amplitude. Thus, none of the time harmonics can pass through the isolator in the backward direction.

Such a contrast between the forward and backward transmissions is based on the asymmetry of the structure, where the two temporal loops possess different modulation amplitudes. Here, the modulation amplitude of the right loop is much larger than the left loop. Hence, the right loop produces much stronger time harmonics and the left loop can be used for fine tuning of the isolator parameters, e.g. the amplitude of the transmitted signals. The half-wavelength long middle interconnector and the quarter-wavelength long arms of the two power splitters are essential parts of this architecture, providing the required destructive and constructive interferences at the fundamental and first higher and lower order time harmonics.

\begin{figure*}
	\centering
	\includegraphics[width=100mm]{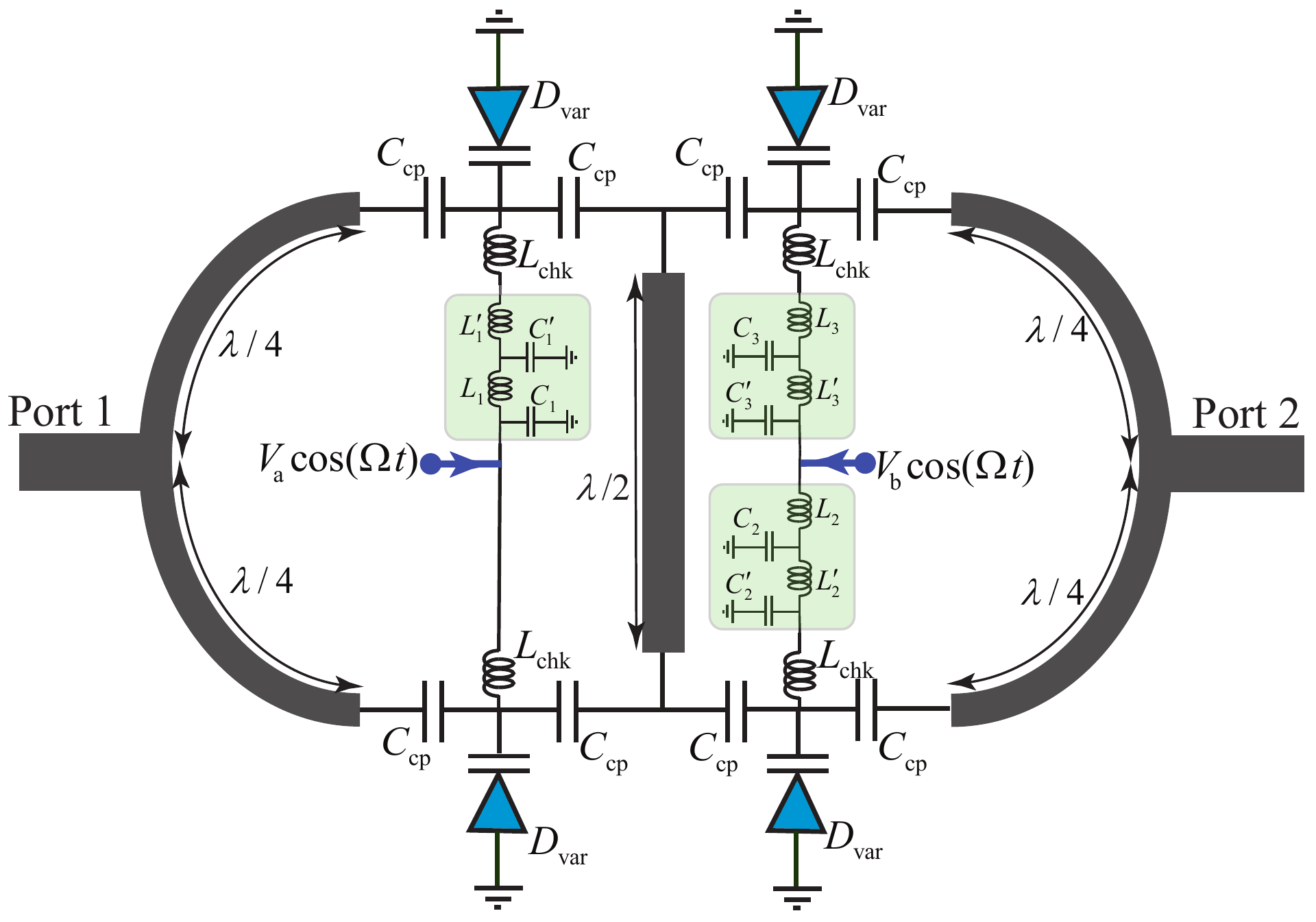}
	\caption{Implementation of the time-modulated isolator by four phase-shifted time-varying varactors.}
	\label{fig:2}
\end{figure*}

Figure~\ref{fig:2} shows the implementation of the time-modulated isolator by four phase-shifted time-varying varactors. Here, four varactors are used for the creation of phase-shifted time modulation. Two power splitters feed the four temporal transmission lines, and eight fixed capacitances are considered for decoupling of the DC bias and low frequency modulation signal from the incident and transmitted high frequency incident microwave signal.

\begin{figure}[h!]
	\begin{center}
		\subfigure[]{\label{fig:ph1} 
			\includegraphics[width=0.4\columnwidth]{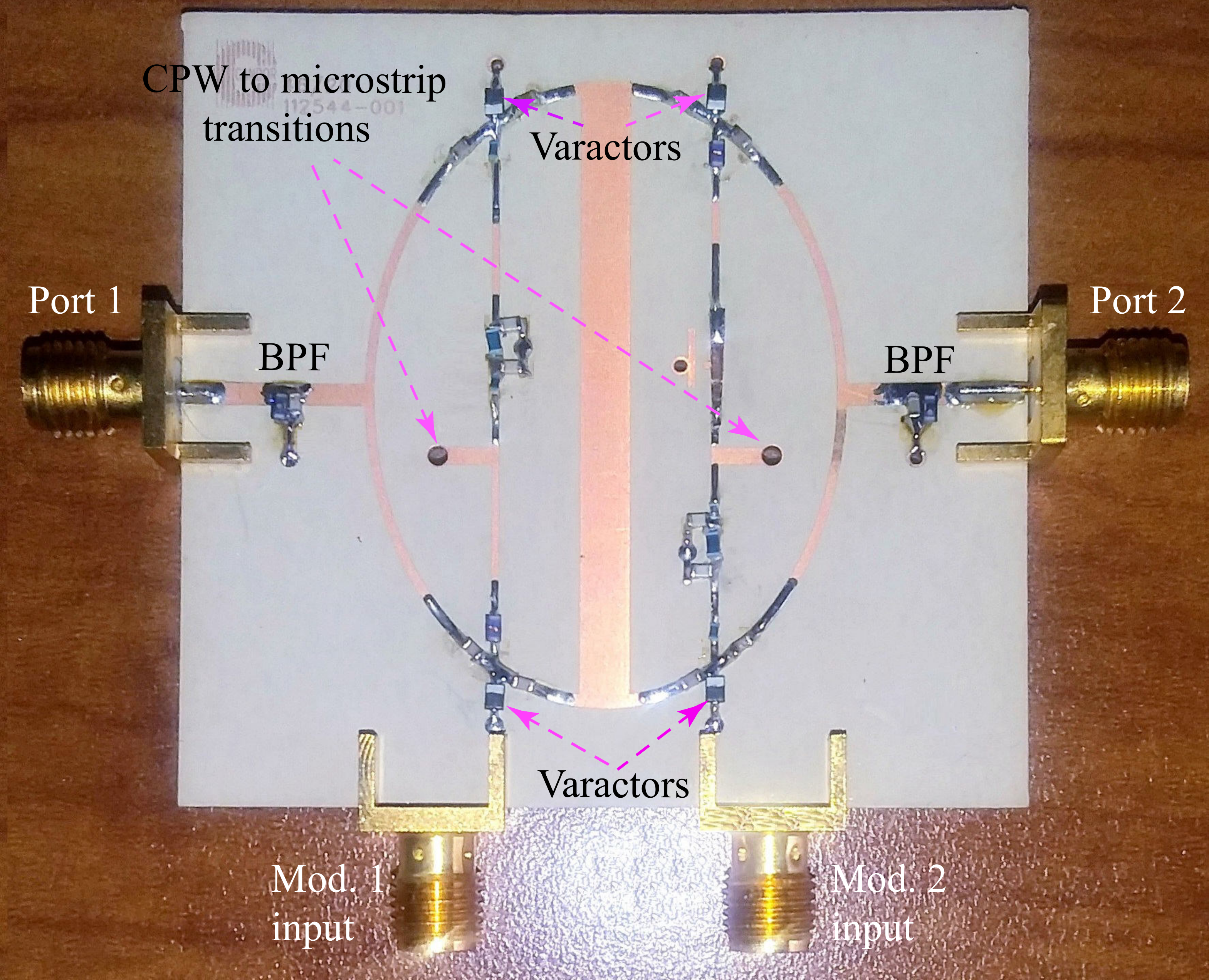}} 
		\subfigure[]{\label{fig:ph2} 
			\includegraphics[width=0.4\columnwidth]{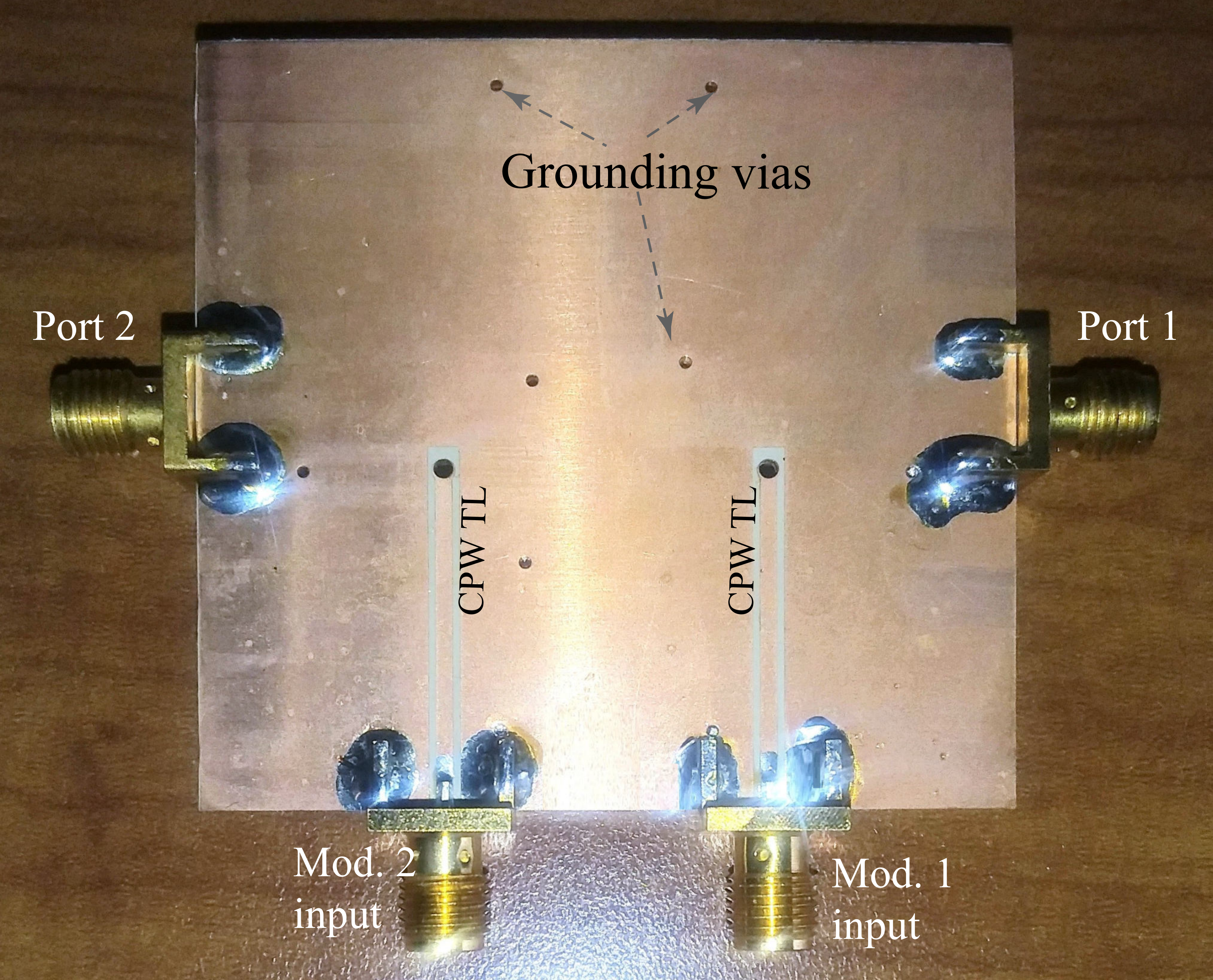}}
		\caption{Photos of the fabricated lightweight low-profile magnetless time-modulated isolator. (a) Top layer. (b) Bottom layer.} 
		\label{Fig:photo}
	\end{center}
\end{figure}

Figures~\ref{fig:ph1} and~\ref{fig:ph2} show photos of the top and bottom layers of the fabricated lightweight low-profile magnetless time-modulated isolator. This isolator is designed at the center frequency of 1.48 GHz with the modulation frequency of $\Omega=2\pi\times103.8$~MHz and modulation power of 20~dBm. Port 1 and Port 2 are the main ports of the isolator for incidence of the microwave signal, while two other ports, i.e., Mod. 1 input and Mod. 2 input, are considered for the modulation signal injection with different phase shifts.

\begin{figure}
	\begin{center}
		\subfigure[]{\label{fig:v0} 
			\includegraphics[width=0.4\columnwidth]{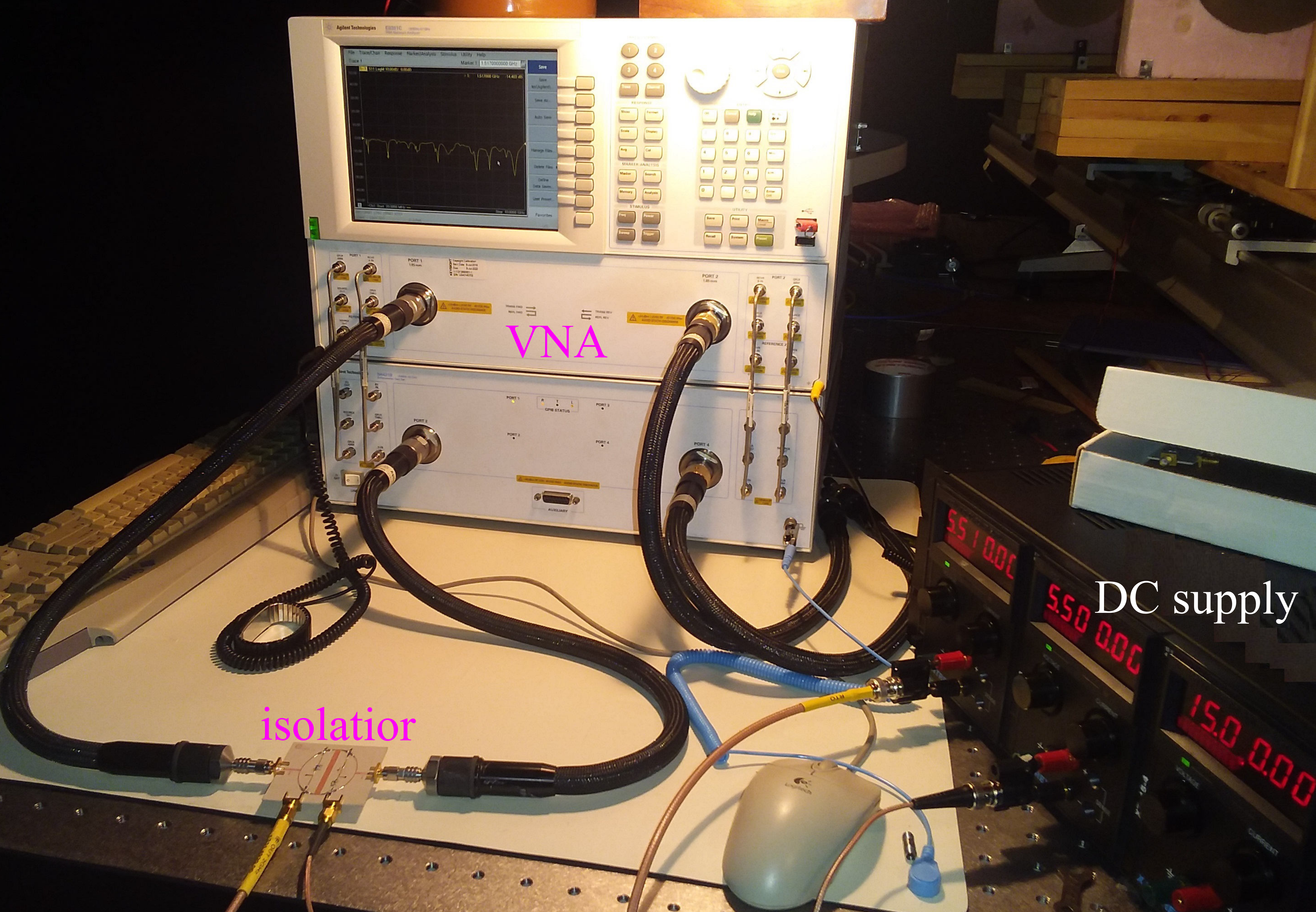}}
		\subfigure[]{\label{fig:v1} 
			\includegraphics[width=0.38\columnwidth]{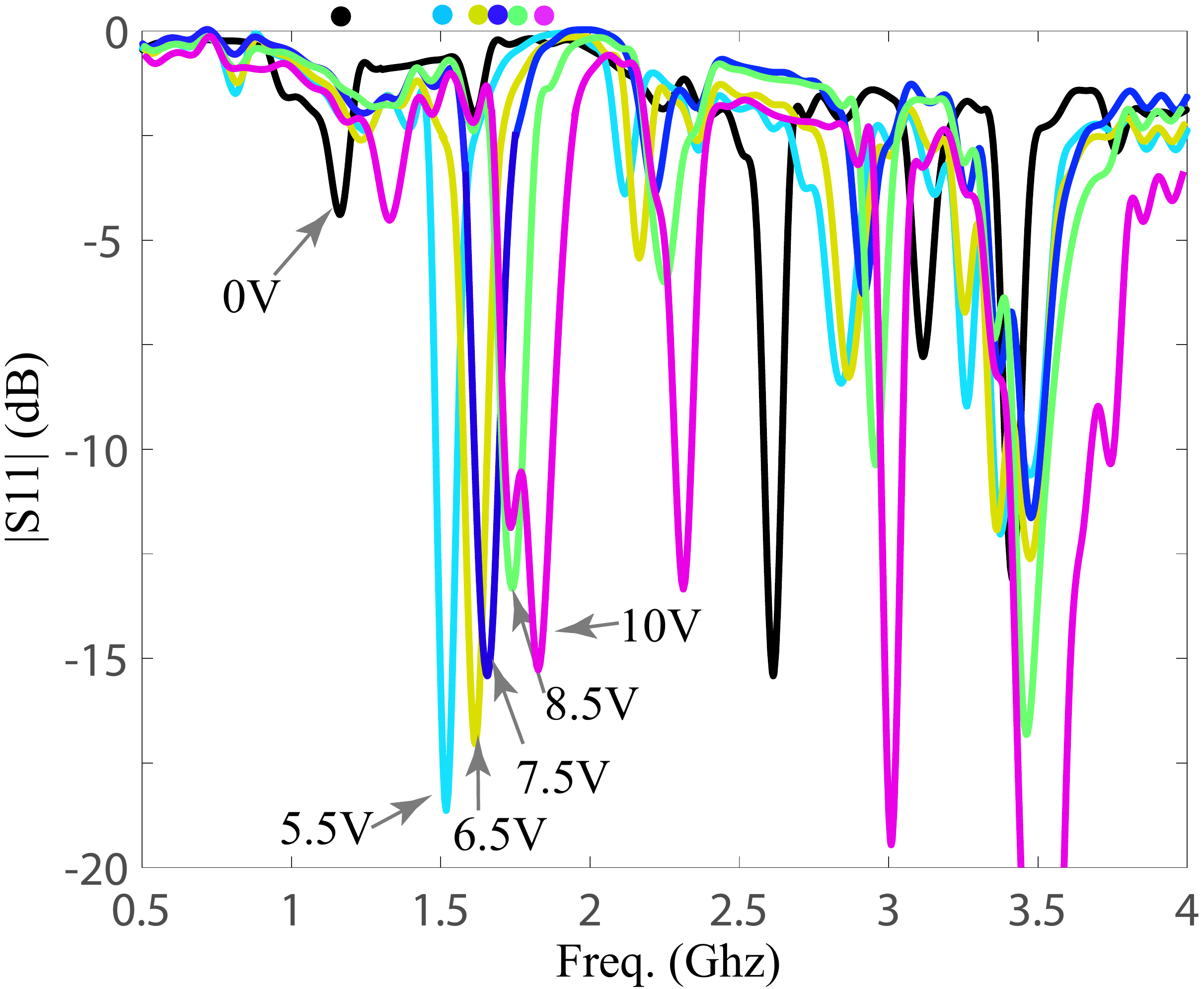}} 
		\subfigure[]{\label{fig:v7} 
			\includegraphics[width=0.38\columnwidth]{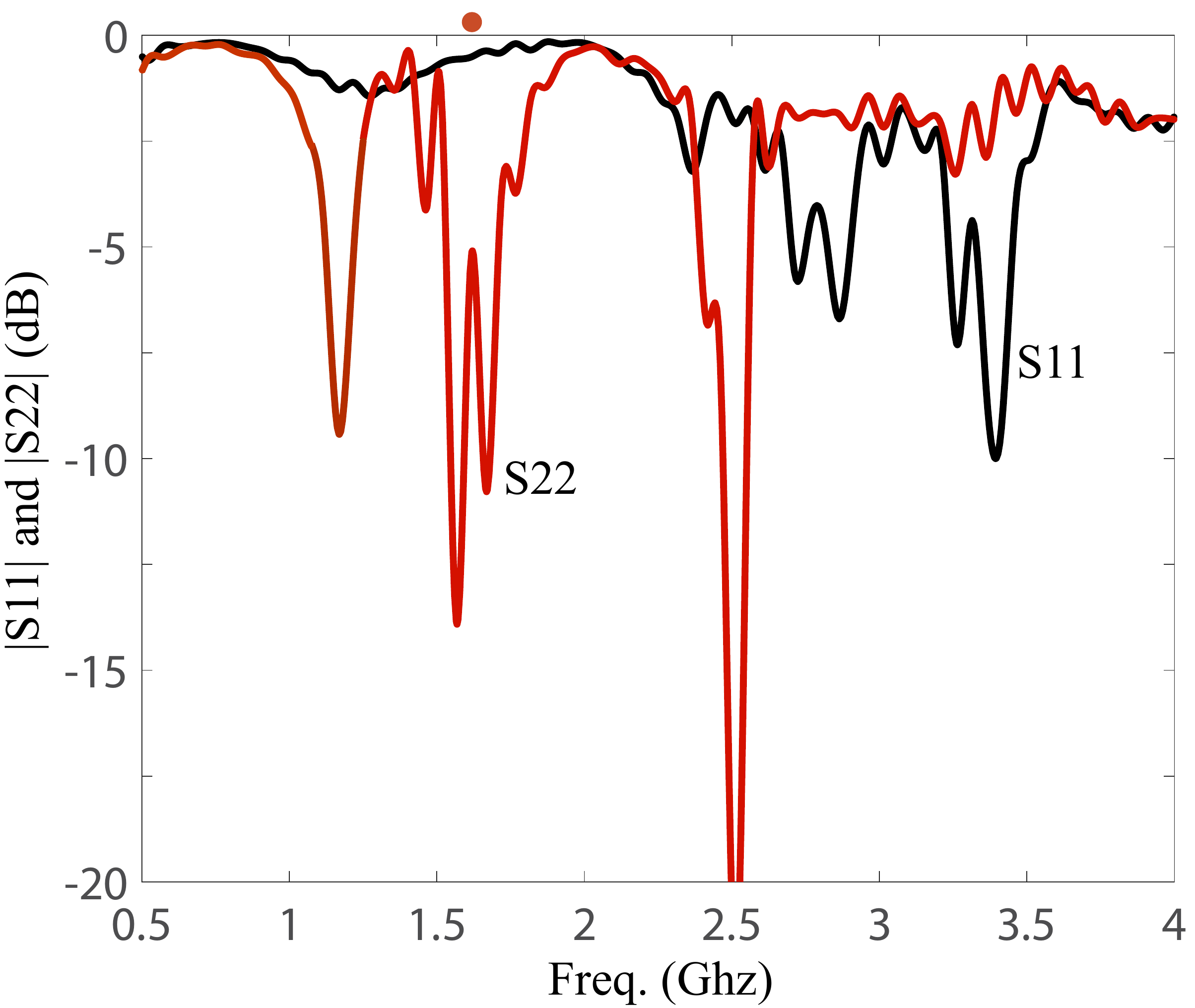}} 
		\subfigure[]{\label{fig:v8} 
			\includegraphics[width=0.38\columnwidth]{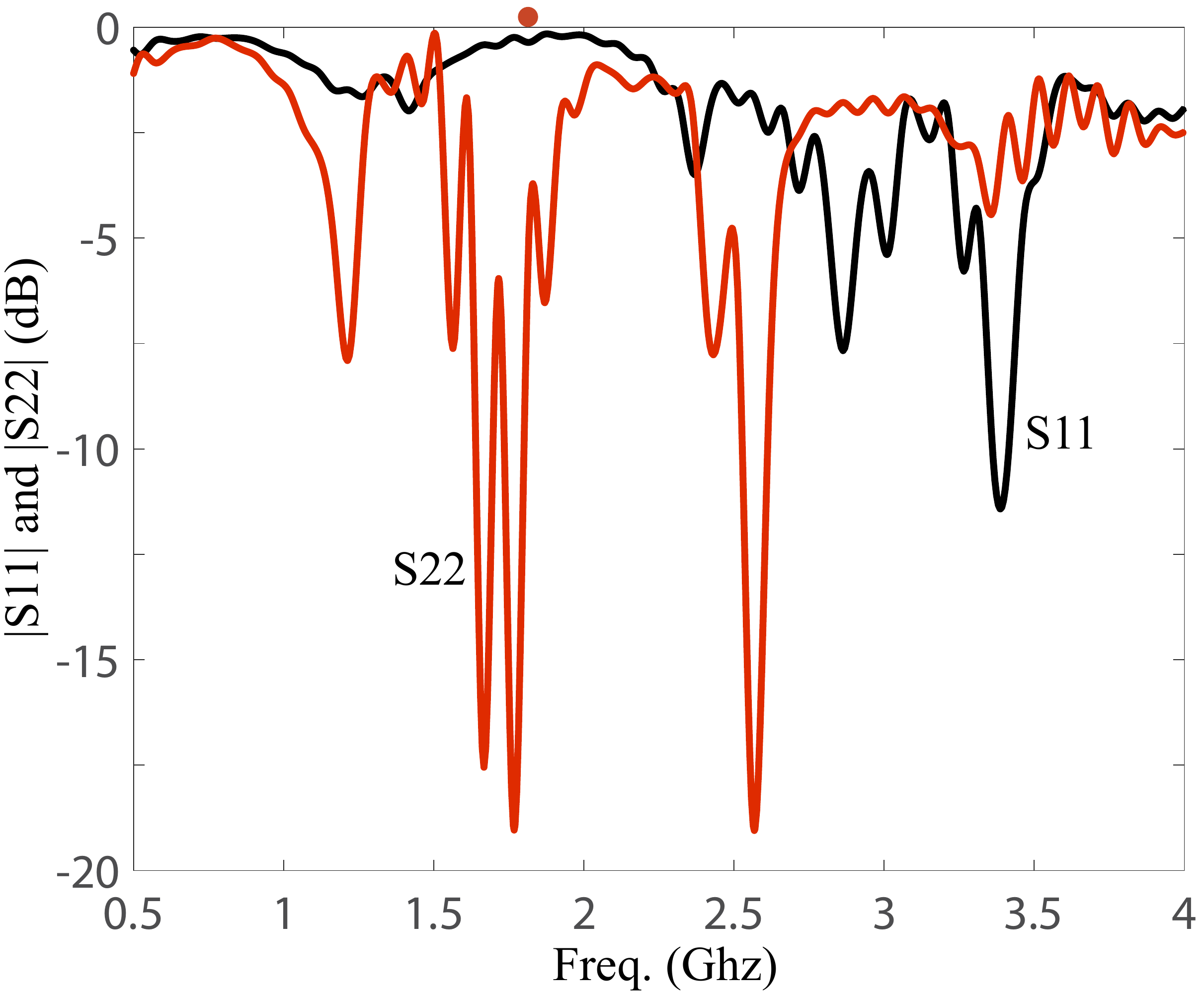}}
		\caption{Experimental results for the input matching of the isolator for different bias voltages when the modulation is off ($\Omega=0$). (a) Measurement setup. (b)~Equal dc voltages on all varactors. (c)~Unequal dc voltage for the left and right varactors, i.e., $V_\text{L}=1V$ and $V_\text{R}=7V$. (d)~Unequal dc voltage for the left and right varactors, i.e., $V_\text{L}=2V$ and $V_\text{R}=9V$. The dots on top of the (b)-(d) plots highlight the variation of the operation frequency by changing the dc bias voltage of the varactor-loaded loops.} 
		\label{Fig:vna}
	\end{center}
\end{figure}

We next demonstrate the experimental results for the scattering parameters of the isolator for different bias voltages when the modulation is off ($\Omega=0$). Figure~\ref{fig:v0} shows the measurement setup for the input matching of the two main ports of the isolator using an E8361C Agilent vector network analyzer. Figure~\ref{fig:v1} plots the input matching of two ports of the isolator for equal dc voltages for all varactors ($|S_{11}|=|S_{22}|$). We next measure the input matching of the two ports of the isolator where the loops are supplied by unequal dc voltages. Figure~\ref{fig:v7} plots the input matching where the left loop is fed by the dc voltage of $V_\text{L}=1V$ and the right loop is fed by the dc voltage of  $V_\text{R}=7V$. It may be seen from this figure that a supplied dc voltage of 7V for the right loop has opened up a desired input matching ($|S_{11}|<-13|$ dB) around 1.5 GHz, whereas a much lower supplied dc voltage of 1V for the left loop has not opened up a desired input matching ($|S_{11}|>-2$ dB). This is due to the fact that the capacitance of the varactors in each loop vary with the supplied voltage and introduces different impedances, i.e., the higher the dc voltage, the lower the capacitance of the varactors. Figure~\ref{fig:v8} plots the input matching for another set of unequal dc voltages for the right and left loops, i.e., $V_\text{L}=2V$ and $V_\text{R}=9V$. Figures~\ref{fig:v1}-~\ref{fig:v8} show that the operation frequency and input matching of the two ports of the isolator can be controlled individually by the supplied dc voltage of the varactors that determines the average capacitance of the varactors. To achieve a desired isolation at a given frequency, we shall ensure that both ports of the isolator are matched through the dc bias voltage of varactors.

\begin{figure}
	\begin{center}
		\subfigure[]{\label{fig:meas_su} 
			\includegraphics[width=80mm]{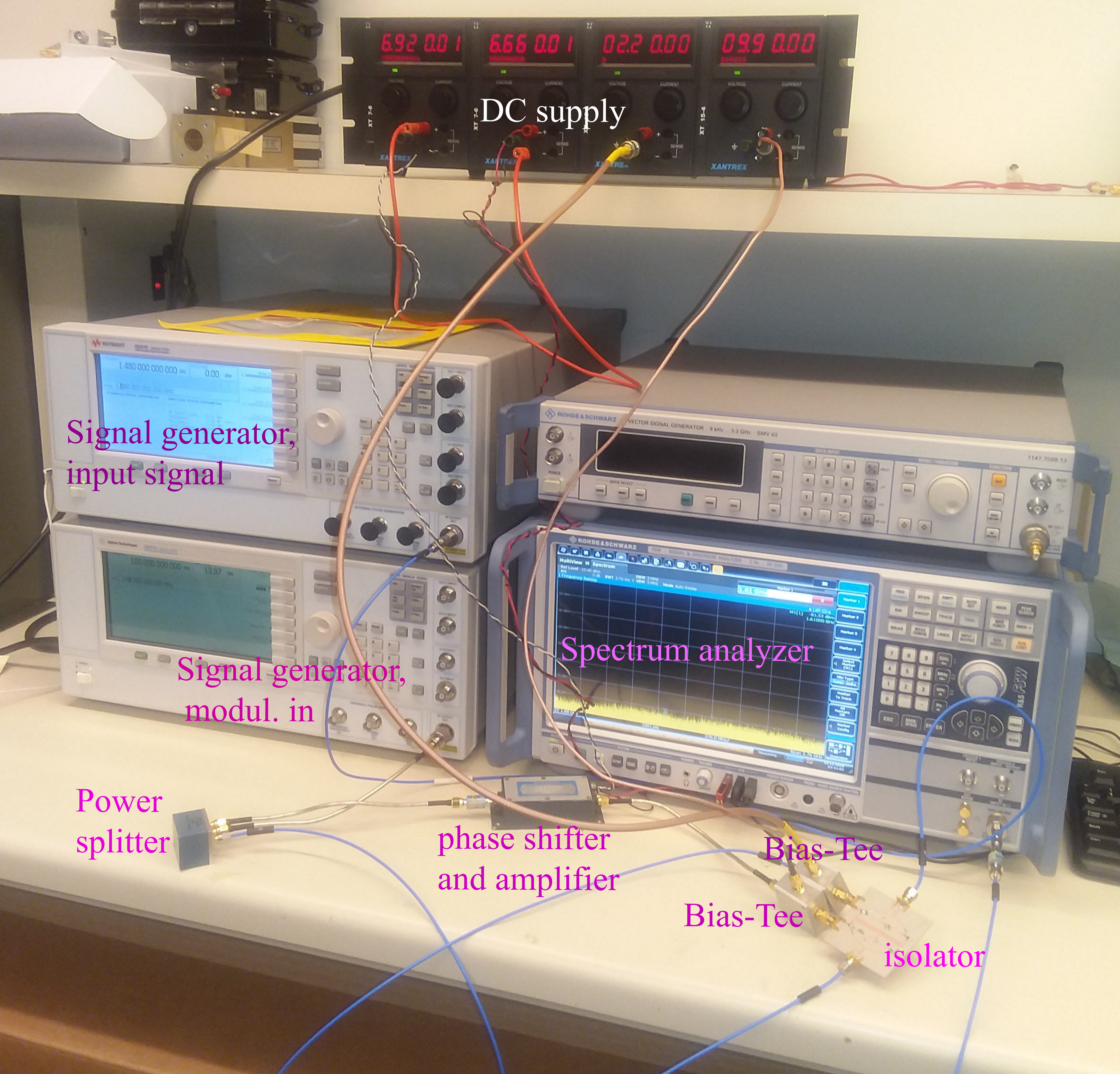}}					
		\subfigure[]{\label{fig:measf} 
			\includegraphics[width=0.8\columnwidth]{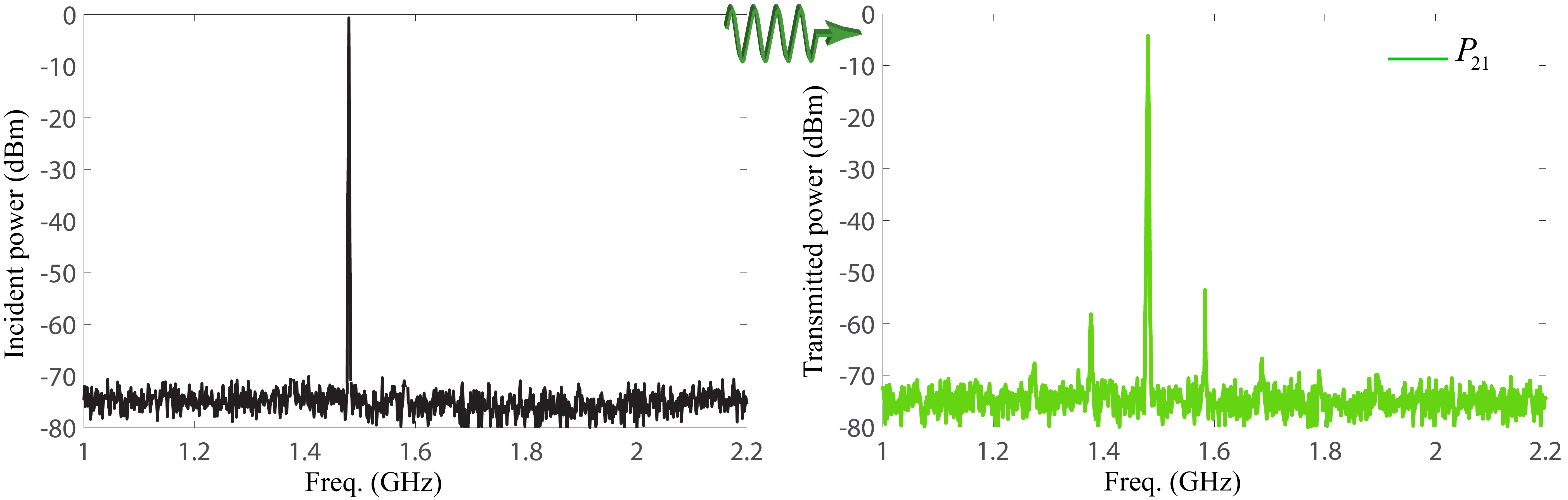}} 
		\subfigure[]{\label{fig:measb} 
			\includegraphics[width=0.8\columnwidth]{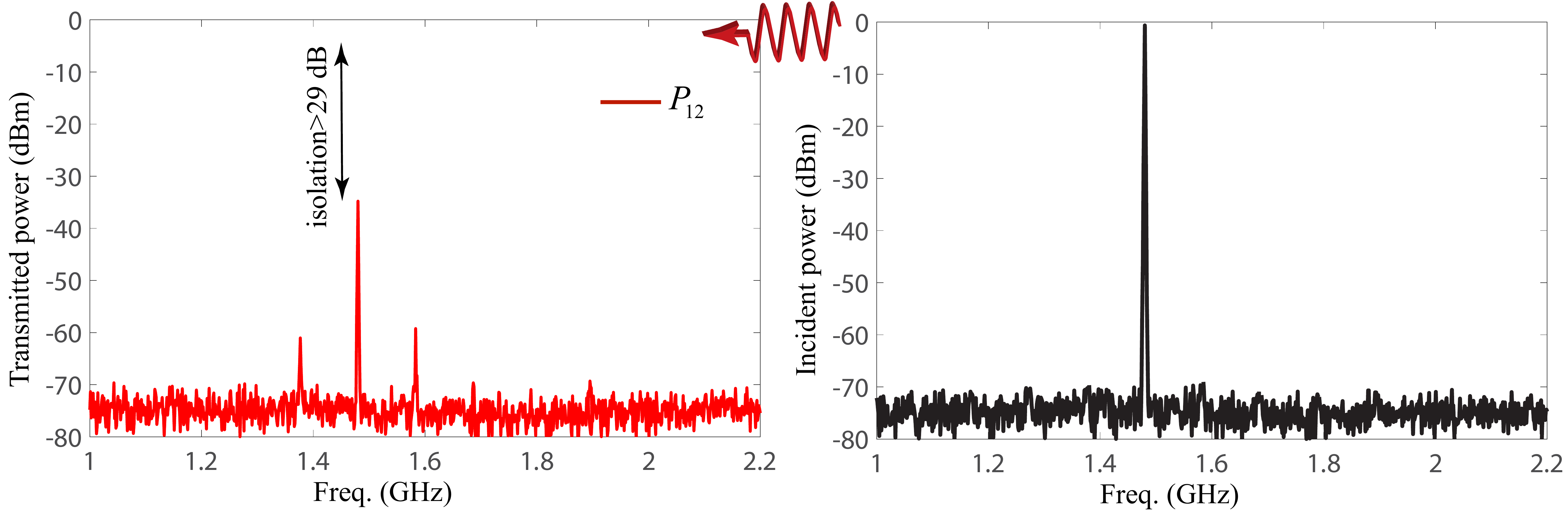}}
		\caption{Experimental demonstration. (a) A photo of the measurement setup. (b) Forward transmission. (c) Backward transmission.} 
		\label{Fig:meas}
	\end{center}
\end{figure}

Figure~\ref{fig:meas_su} shows a photo of the experimental set-up for the measurement of the unidirectional signal transmission through the temporal isolator, where $\Omega=2\pi\times103.8$~MHz. Figure~\ref{fig:measf} plots the experimental results for forward wave incidence and transmission, while Fig.~\ref{fig:measb} plots the experimental results for backward wave incidence and transmission. It may be seen from the experimental results that the contrast between the forward and backward transmitted waves is more than 27 dB, whereas the undesired time harmonics are highly suppressed.

Table~\ref{Tab:1} summarizes the performance of the fabricated prototype and compares its performance with those of recently proposed isolators. In this table, NF stands for noise figure, IL stands for insertion loss, RL stands for return loss, Isol. denotes the isolation level, PH stands for power handling, and FBW is an acronym for the fractional bandwidth which is equal to the frequency bandwidth over the center frequency. Additionally, OP$_\text{1dB}$ is the output 1 dB compression point which shows the prominent linearity of the isolator. Table~\ref{Tab:1} lists isolators realized through different technologies, i.e., space-time (ST) modulation, magnetic-ferrite-loaded waveguides, nonlinearity, and transistor-loaded transmission lines. The proposed isolator can be realized in a much more compact fashion, and may be integrated into chip technology at high frequencies thanks to the availability of variable capacitors at such frequencies~\cite{lira2012electrically}. In addition, in contrast to transistor-based isolators~\cite{lee20055,Wang_TMTT_2015} where nonlinear and low power handling (PH) transistors are placed in series with the incident signal, our isolator is endowed by high power handling as the varactors are placed in parallel to the incident signal. In our experiment, we applied up to +33~dBm input signal, and the power handling is expected to exceed $47$~dBm.

Additionally, the proposed time-modulated isolator exhibits a much lower noise figure of 3.4 dB in comparison to two transistor-based isolators, 13 dB NF of Ref.~\cite{lee20055} and 6.7 dB NF of Ref.~\cite{Wang_TMTT_2015}. It can be argued that one of the main reasons that transistor-based circulators never became main stream is their poor noise performance. However, varactor diodes introduce much lower noise to the system~\cite{van1970noise}. There exist several potential sources of noise, that is, thermal noise of the series resistance of a varactor, shot noise of the forward current, shot noise of the reverse current, ionisation noise and reverse-breakdown noise~\cite{hyde1964varactor}. Nevertheless, it has been proved, theoretically and experimentally, that varactor-based parametric amplifiers and circuits introduce much lower noise in comparison to other techniques~\cite{kita1963low,van1970noise,lee2010low,li2018nonreciprocal}.

\begin{figure}
\begin{center}
	\centering
	\includegraphics[width=0.5\columnwidth]{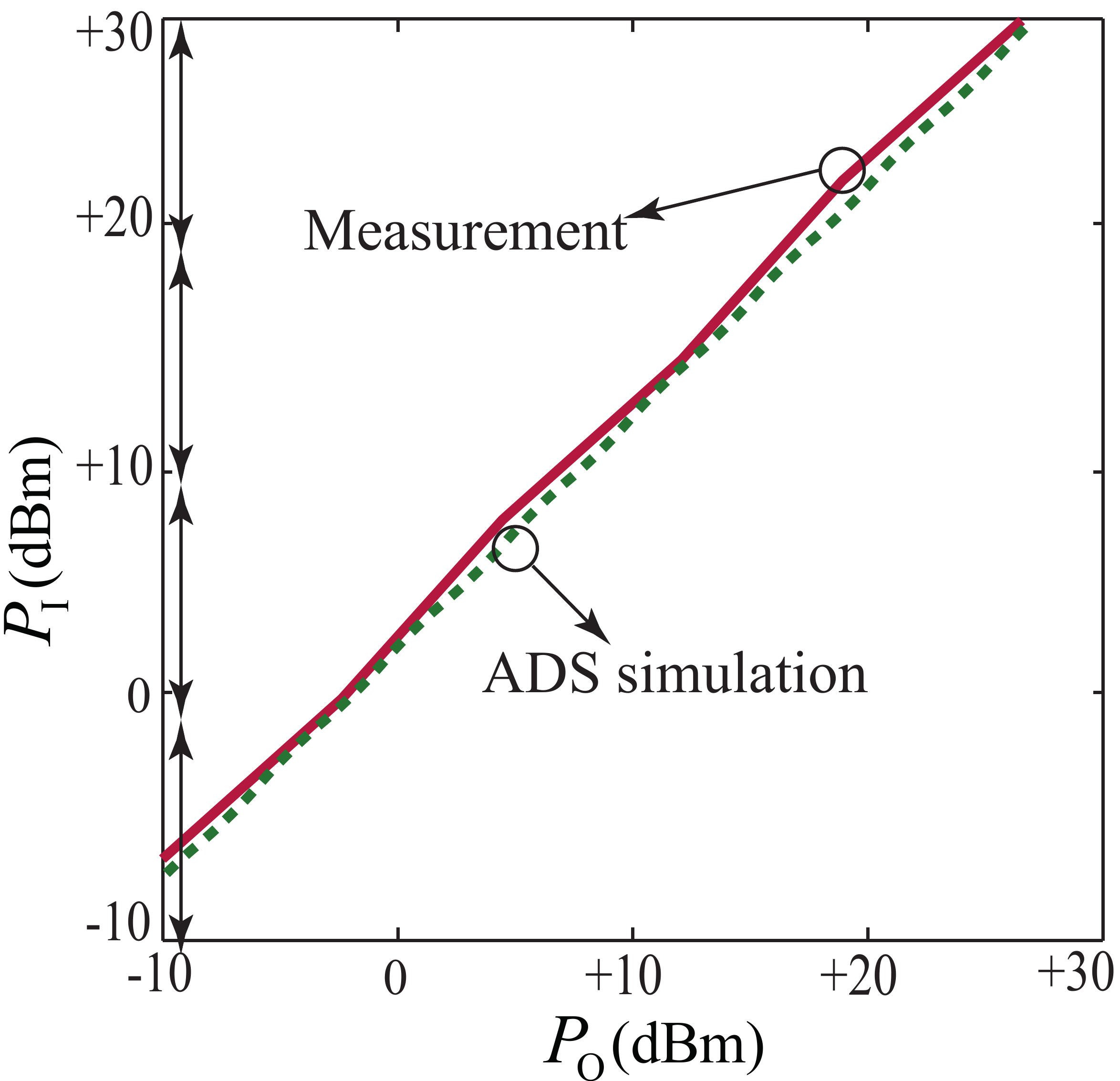}
	\caption{Simulation and experimental results for linearity of the proposed temporal isolator.}
	\label{Fig:sim}
\end{center}
\end{figure}

By leveraging the linear response of the time modulation, the proposed temporal isolator exhibits an outstanding linear response, where the OP$_\text{1dB}$ is more than 31 dBm. Figure~\ref{Fig:sim} plots the simulation (using Keysight's Advanced Design System commercial software) and experimental results for the output power $P_\text{O}$ versus input power $P_\text{I}$ which shows the output signal follows the input signal linearly.

\section{Conclusions}

In this study, we have successfully demonstrated a phase-engineered temporal-loop-based isolator. This isolator is featuring large isolation levels, weak undesired time harmonics and a low profile. The experimental demonstration of the proposed time-modulated isolator is carried out at microwave frequencies. The experimental results show strong unidirectional wave transmission through the isolator with more than 27 dB contrast between the forward and backward waves across a fractional bandwidth of $14.3\%$. Such an isolator outperforms the nonlinear-based and transistor-based isolators by featuring a highly linear response with OP$_\text{1dB}$ of higher than $31$~dBm, high power rating of more than $47$~dBm, and a low noise figure of $3.4$~dB.


\section{Experimental Section}

This isolator is designed at the center frequency of 1.48 GHz with the modulation frequency of $\Omega=2\pi\times103.8$ MHz and modulation power of 20 dBm. The isolator is implemented on a RT6010 substrate with permittivity $\epsilon_\text{r} = 10.2$, thickness $h = 50$~mil and $\tan \delta = 0.0023$. The two 50 Ohm coplanar waveguide (CPW) transmission lines at the back of the structure possess a width of W=45 mils and a gap of G=20 mils. We have used four BB837 varactors manufactured by Infineon Technologies for realizing the phase-shifted time modulation. We utilized two ZFBT-6GQ+ minicircuits bias-tees.

\medskip
\textbf{Supporting Information} \par 
Supporting Information is available from the Wiley Online Library or from the author.

\medskip

\textbf{Acknowledgements}\par 
This work is supported by the Natural Sciences and Engineering Research Council of Canada (NSERC).

\medskip

\textbf{Conflict of Interest}\par 
The authors declare no conflict of interests.

\medskip

\textbf{Author contributions}\par 
S.T. carried out the analytical modeling, numerical simulations, sample fabrication, and measurements. G.V.E. planned, coordinated, and supervised the work. All authors discussed the theoretical and experimental aspects and interpreted the results. All authors contributed to the preparation and writing of the manuscript. Correspondence and requests for materials should be addressed to Sajjad Taravati~(email: sajjad.taravati@utoronto.ca).

\medskip

\textbf{Data Availability}\par The data that supports the findings of this study are available in the supplementary material of this article.

%


{
	\setlength{\tabcolsep}{1.25mm}%
	\renewcommand{\arraystretch}{1.6}
	\newcommand{\CPcolumnonewidth}{not used}%
	\newcommand{\CPcolumntwowidth}{23mm}%
	\newcommand{\CPcolumnthreewidth}{12mm}%
	\newcommand{\CPcolumnfourwidth}{33mm}%
	\begin{table*}[h!]
		\caption{Summary of the efficiency of the time-modulated isolator and comparison with previously reported isolators.}
		\medskip
		\centering
		\begin{tabular}{|>{\centering\arraybackslash}m{1.2cm}|>{\centering\arraybackslash}m{2.4cm}| >{\centering\arraybackslash}m{3.2cm}|>{\centering\arraybackslash}m{1cm}|>{\centering\arraybackslash}m{1cm}|>{\centering\arraybackslash}m{1.2cm}|>{\centering\arraybackslash}m{1.2cm}|>{\centering\arraybackslash}m{1.2cm}|>{\centering\arraybackslash}m{1cm}|>{\centering\arraybackslash}m{2.5cm}|}
			\hline
			Isolator  & Technology&  Size & \shortstack{NF\\(dB) } & \shortstack{FBW\\($\%$)}& \shortstack{Isol.\\ (dB)}& \shortstack{IL\\(dB) }&\shortstack{RL\\(dB)}&\shortstack{ OP$_\text{1dB}$\\ (dBm)} & Properties\\
			\hline
			This work & Time modulation & \shortstack{Small, $<<\lambda_\text{c}^2$\\ ($53\times55~\text{mm}^2$\\ at 1.48 GHz)} & 3.4 & 14.2  & $>27$  & $<3.4$ & $>16.3$& $>31$ & \shortstack{lightweight,\\low-profile,\\low noise,\\linear,\\high PH\\RF bias}\\ \hline
			Ref.~\cite{Taravati_PRB_2017}  & Space-time modulation & \shortstack{Moderate, $\approx \lambda_\text{c}^2/2$\\ ($117\times56~\text{mm}^2$\\ at 1.5 GHz)}& 4.8& 6 & 20 & $<2$ & $>30$ & $>21$ &\shortstack{linear,\\Large profile,\\RF bias} \\\hline
			Ref.~\cite{Taravati_PRB_SB_2017}  & \shortstack{Space-time\\ modulation\\ (self-biased)} & \shortstack{Moderate, $\approx \lambda_\text{c}^2$\\ ($184\times62~\text{mm}^2$\\ at 2 GHz)}& 4.6 & 7.4 & 25.4  & $<2$ & $>25$ & $>22$ & \shortstack{Moder. FBW,\\No RF bias,\\ linear}\\\hline
			Ref.~\cite{seewald2010ferrite} & \shortstack{Ferrite-loaded\\ waveguide}& \shortstack{Moderate, $\approx \lambda_\text{c}^2/2$\\ ($30\times6~\text{mm}^2$\\ at 15.4 GHz)}& N/A &  19 & 22  & 4.6 & N/A &  N/A & \shortstack{Cumbersome,\\ magnetic bias}\\\hline				
			Ref.~\cite{Hewlett_Packard}  & \shortstack{Ferrite-loaded\\ waveguide} & \shortstack{Large, $>>\lambda_\text{c}^2$\\ ($83\times33~\text{mm}^2$\\ at 33.25 GHz)} & N/A & 40 &25 & 1.5 & 15& 33 &\shortstack{Heavy,\\cumbersome}\\\hline
			Ref.~\cite{Millitech} & \shortstack{Ferrite-loaded\\ waveguide} &\shortstack{Large, $>>\lambda_\text{c}^2$\\ ($81\times33~\text{mm}^2$\\ at 33.25 GHz)}&N/A &  40 & 27 & 1.5 & 8 & 34& \shortstack{Heavy and \\cumbersome}\\\hline	
			Ref.~\cite{sounas2018broadband}  & Nonlinearity & \shortstack{Moderate, $\approx \lambda_\text{c}^2$\\  at 0.75 GHz} &N/A & 5.33 & 25 & $<12$ & N/A & N/A & Narrow-band\\\hline	
			Ref.~\cite{lee20055}  & \shortstack{Transistor\\ (200-nm HBT)} &\shortstack{Small, $<<\lambda_\text{c}^2$\\ ($0.64\times0.43~\text{mm}^2$\\ at 33.25 GHz)} & 13 &18 & 30 & 2 & $>10$& N/A& \shortstack{Very low PH,\\ high NF} \\
			\hline
			Ref.~\cite{Wang_TMTT_2015}  & \shortstack{Transistor\\ (180-nm\\CMOS)} & \shortstack{Small, $<<\lambda_\text{c}^2$\\ ($0.66\times0.435~\text{mm}^2$\\ at 24 GHz)}& 6.7& 1 & 36.8 & 1.8 & $>15$ & $>13$& \shortstack{Very low PH,\\ high NF,\\ narrow-band}\\
			\hline
		\end{tabular}
		\label{Tab:1}
	\end{table*}
}

\pagebreak

\newpage
\bibliographystyle{ieeetran}
\bibliography{Taravati_Reference}

\begin{thebibliography}{10}
\providecommand{\url}[1]{#1}
\csname url@samestyle\endcsname
\providecommand{\newblock}{\relax}
\providecommand{\bibinfo}[2]{#2}
\providecommand{\BIBentrySTDinterwordspacing}{\spaceskip=0pt\relax}
\providecommand{\BIBentryALTinterwordstretchfactor}{4}
\providecommand{\BIBentryALTinterwordspacing}{\spaceskip=\fontdimen2\font plus
\BIBentryALTinterwordstretchfactor\fontdimen3\font minus
  \fontdimen4\font\relax}
\providecommand{\BIBforeignlanguage}[2]{{%
\expandafter\ifx\csname l@#1\endcsname\relax
\typeout{** WARNING: IEEEtran.bst: No hyphenation pattern has been}%
\typeout{** loaded for the language `#1'. Using the pattern for}%
\typeout{** the default language instead.}%
\else
\language=\csname l@#1\endcsname
\fi
#2}}
\providecommand{\BIBdecl}{\relax}
\BIBdecl

\bibitem{SOOHOO_TM_1968}
R.~F. Soohoo, ``Microwave ferrite materials and devices,'' \emph{{IEEE Trans.
  Magn.}}, vol.~4, no.~2, pp. 118--133, Jun. 1968.

\bibitem{nagulu2018nonreciprocal}
A.~Nagulu, T.~Dinc, Z.~Xiao, M.~Tymchenko, D.~L. Sounas, A.~Al{\`u}, and
  H.~Krishnaswamy, ``Nonreciprocal components based on switched transmission
  lines,'' \emph{IEEE Transactions on Microwave Theory and Techniques},
  vol.~66, no.~11, pp. 4706--4725, 2018.

\bibitem{Taravati_Kishk_MicMag_2019}
S.~Taravati and A.~A. Kishk, ``Space-time modulation: Principles and
  applications,'' \emph{IEEE Microw. Mag.}, vol.~21, no.~4, pp. 30--56, 2020.

\bibitem{Hirota_TED_1971}
K.~Suzuki and R.~Hirota, ``Nonreciprocal millimeter-wave devices using a
  solid-state plasma at room temperature,'' \emph{IEEE Trans. Electron
  Devices}, vol.~18, no.~7, pp. 408--411, 1971.

\bibitem{jawad2017millimeter}
G.~N. Jawad, C.~I. Duff, and R.~Sloan, ``A millimeter-wave gyroelectric
  waveguide isolator,'' \emph{{IEEE Trans. Microw. Theory Techn.}}, vol.~65,
  no.~4, pp. 1249--1256, 2017.

\bibitem{lee20055}
J.~Lee, J.~D. Cressler, and A.~J. Joseph, ``A 5-6 {GHz} {S}i{G}e {HBT}
  monolithic active isolator for improving reverse isolation in wireless
  systems,'' \emph{{IEEE Microw. Wireless Compon. Lett.}}, vol.~15, no.~4, pp.
  220--222, 2005.

\bibitem{chang2015design}
J.-F. Chang, J.-C. Kao, Y.-H. Lin, and H.~Wang, ``Design and analysis of
  24-{GHz} active isolator and quasi-circulator,'' \emph{{IEEE Trans. Microw.
  Theory Techn.}}, vol.~63, no.~8, pp. 2638--2649, 2015.

\bibitem{Taravati_2017_NR_Nongyro}
S.~Taravati, B.~A. Khan, S.~Gupta, K.~Achouri, and C.~Caloz, ``Nonreciprocal
  nongyrotropic magnetless metasurface,'' \emph{{IEEE Trans. Antennas
  Propagat.}}, vol.~65, no.~7, pp. 3589--3597, Aug. 2017.

\bibitem{wang2019highly}
Y.~Wang, W.~Chen, and X.~Chen, ``Highly linear and magnetless isolator based on
  weakly coupled nonreciprocal metamaterials,'' \emph{{IEEE Trans. Microw.
  Theory Techn.}}, vol.~67, no.~11, pp. 4322--4331, 2019.

\bibitem{taravati2021programmable}
S.~Taravati and G.~V. Eleftheriades, ``Programmable nonreciprocal meta-prism,''
  \emph{Sci. Rep.}, vol.~11, no.~1, pp. 1--12, 2021.

\bibitem{wang2020210}
Y.~Wang, W.~Chen, and X.~Li, ``A 210-{GHz} magnetless nonreciprocal isolator in
  130-nm sige bicmos based on resistor-free unidirectional ring resonators,''
  \emph{IEEE Microwave and Wireless Components Letters}, vol.~30, no.~5, pp.
  524--527, 2020.

\bibitem{taravati2021full}
S.~Taravati and G.~V. Eleftheriades, ``Full-duplex reflective beamsteering
  metasurface featuring magnetless nonreciprocal amplification,'' \emph{arXiv
  preprint arXiv:2101.10067}, 2021.

\bibitem{tsutsumi1987dielectric}
M.~Tsutsumi, K.~Tanaka, and N.~Kumagai, ``Dielectric slab waveguide isolator in
  the millimeter wave frequency,'' \emph{IEEE Trans. Magn.}, vol.~23, no.~5,
  pp. 3739--3741, 1987.

\bibitem{seewald2010ferrite}
C.~K. Seewald and J.~R. Bray, ``Ferrite-filled antisymmetrically biased
  rectangular waveguide isolator using magnetostatic surface wave modes,''
  \emph{{IEEE Trans. Microw. Theory Techn.}}, vol.~58, no.~6, pp. 1493--1501,
  2010.

\bibitem{wu2012novel}
J.~Wu, M.~Li, X.~Yang, S.~Beguhn, and N.~X. Sun, ``A novel tunable planar
  isolator with serrated microstrip structure,'' \emph{IEEE Trans. Magn.},
  vol.~48, no.~11, pp. 4371--4374, 2012.

\bibitem{cheng2014narrowband}
Y.~J. Cheng, Q.~D. Huang, Y.~R. Wang, and J.~L.-W. Li, ``Narrowband substrate
  integrated waveguide isolators,'' \emph{{IEEE Microw. Wireless Compon.
  Lett.}}, vol.~24, no.~10, pp. 698--700, 2014.

\bibitem{farooqui2014inkjet}
M.~F. Farooqui, A.~Nafe, and A.~Shamim, ``Inkjet printed ferrite-filled
  rectangular waveguide x-band isolator,'' in \emph{2014 IEEE MTT-S
  International Microwave Symposium (IMS2014)}.\hskip 1em plus 0.5em minus
  0.4em\relax IEEE, 2014, pp. 1--4.

\bibitem{marynowski2018integrated}
W.~Marynowski, ``Integrated broadband edge-guided mode isolator with
  antiparallel biasing of the ferrite slabs,'' \emph{{IEEE Microw. Wireless
  Compon. Lett.}}, vol.~28, no.~5, pp. 392--394, 2018.

\bibitem{ghaffar2019theory}
F.~A. Ghaffar, J.~R. Bray, M.~Vaseem, L.~Roy, and A.~Shamim, ``Theory and
  design of tunable full-mode and half-mode ferrite waveguide isolators,''
  \emph{IEEE Trans. Magn.}, vol.~55, no.~8, pp. 1--8, 2019.

\bibitem{sounas2018broadband}
D.~L. Sounas, J.~Soric, and A.~Alu, ``Broadband passive isolators based on
  coupled nonlinear resonances,'' \emph{Nat. Electronics}, vol.~1, no.~2, pp.
  113--119, 2018.

\bibitem{Fan_PRL_109_2012}
H.~Lira, Z.~Yu, S.~Fan, and M.~Lipson, ``Electrically driven nonreciprocity
  induced by interband photonic transition on a silicon chip,'' \emph{{Phys.
  Rev. Lett.}}, vol. 109, p. 033901, Jul. 2012.

\bibitem{Wang_TMTT_2014}
S.~Qin, Q.~Xu, and Y.~E. Wang, ``Nonreciprocal components with distributedly
  modulated capacitors,'' \emph{{IEEE Trans. Microw. Theory Techn.}}, vol.~62,
  no.~10, pp. 2260--2272, Oct. 2014.

\bibitem{Taravati_PRB_2017}
S.~Taravati, N.~Chamanara, and C.~Caloz, ``Nonreciprocal electromagnetic
  scattering from a periodically space-time modulated slab and application to a
  quasisonic isolator,'' \emph{{Phys. Rev. B}}, vol.~96, no.~16, p. 165144,
  Oct. 2017.

\bibitem{Taravati_PRB_SB_2017}
S.~Taravati, ``Self-biased broadband magnet-free linear isolator based on
  one-way space-time coherency,'' \emph{{Phys. Rev. B}}, vol.~96, no.~23, p.
  235150, Dec. 2017.

\bibitem{Lax_1962}
B.~Lax and K.~J. Button, \emph{Microwave Ferrites and Ferrimagnetics}.\hskip
  1em plus 0.5em minus 0.4em\relax New York: McGraw-Hill, 1962.

\bibitem{shi2015limitationsNL}
Y.~Shi, Z.~Yu, and S.~Fan, ``Limitations of nonlinear optical isolators due to
  dynamic reciprocity,'' \emph{Nat. Photon.}, vol.~9, no.~6, pp. 388--392,
  2015.

\bibitem{fang2012photonic}
K.~Fang, Z.~Yu, and S.~Fan, ``Photonic aharonov-bohm effect based on dynamic
  modulation,'' \emph{{Phys. Rev. Lett.}}, vol. 108, no.~15, p. 153901, 2012.

\bibitem{estep2014magnetic}
N.~A. Estep, D.~L. Sounas, J.~Soric, and A.~Al{\`u}, ``Magnetic-free
  non-reciprocity and isolation based on parametrically modulated
  coupled-resonator loops,'' \emph{Nat. Phys.}, vol.~10, no.~12, pp. 923--927,
  2014.

\bibitem{taravati_PRApp_2019}
S.~Taravati and G.~V. Eleftheriades, ``Generalized space-time periodic
  diffraction gratings: Theory and applications,'' \emph{{Phys. Rev. Appl.}},
  vol.~12, no.~2, p. 024026, 2019.

\bibitem{ramaccia2021temporal}
D.~Ramaccia, A.~Al{\`u}, A.~Toscano, and F.~Bilotti, ``Temporal multilayer
  structures for designing higher-order transfer functions using time-varying
  metamaterials,'' \emph{Appl. Phys. Lett.}, vol. 118, no.~10, p. 101901, 2021.

\bibitem{elnaggar2020modeling}
S.~Y. Elnaggar and G.~N. Milford, ``Modeling space--time periodic structures
  with arbitrary unit cells using time periodic circuit theory,'' \emph{{IEEE
  Trans. Antennas Propagat.}}, vol.~68, no.~9, pp. 6636--6645, 2020.

\bibitem{Engheta_PRB_2021}
D.~M. Sol\'{\i}s and N.~Engheta, ``Functional analysis of the polarization
  response in linear time-varying media:a generalization of the kramers-kronig
  relations,'' \emph{{Phys. Rev. B}}, vol. 103, no.~14, p. 144303, 2003.

\bibitem{taravati2021pure}
S.~Taravati and G.~V. Eleftheriades, ``Pure and linear frequency-conversion
  temporal metasurface,'' \emph{{Phys. Rev. Appl.}}, vol.~15, no.~6, p. 064011,
  2021.

\bibitem{tien1958traveling}
P.~Tien and H.~Suhl, ``A traveling-wave ferromagnetic amplifier,'' \emph{Proc.
  IEEE}, vol.~46, no.~4, pp. 700--706, 1958.

\bibitem{tien1958parametric}
P.~Tien, ``Parametric amplification and frequency mixing in propagating
  circuits,'' \emph{J. Appl. Phys.}, vol.~29, no.~9, pp. 1347--1357, 1958.

\bibitem{zhu2020tunable}
X.~Zhu, J.~Li, C.~Shen, G.~Zhang, S.~A. Cummer, and L.~Li, ``Tunable
  unidirectional compact acoustic amplifier via space-time modulated
  membranes,'' \emph{{Phys. Rev. B}}, vol. 102, no.~2, p. 024309, 2020.

\bibitem{Taravati_Kishk_PRB_2018}
S.~Taravati and A.~A. Kishk, ``Dynamic modulation yields one-way beam
  splitting,'' \emph{{Phys. Rev. B}}, vol.~99, no.~7, p. 075101, Jan. 2019.

\bibitem{Taravati_PRB_Mixer_2018}
S.~Taravati, ``Aperiodic space-time modulation for pure frequency mixing,''
  \emph{{Phys. Rev. B}}, vol.~97, no.~11, p. 115131, 2018.

\bibitem{zang2019nonreciprocal_metas}
J.~W. Zang, D.~Correas-Serrano, J.~T.~S. Do, X.~Liu, A.~Alvarez-Melcon, and
  J.~S. Gomez-Diaz, ``Nonreciprocal wavefront engineering with time-modulated
  gradient metasurfaces,'' \emph{{Phys. Rev. Appl.}}, vol.~11, no.~5, p.
  054054, 2019.

\bibitem{taravati2020full}
S.~Taravati and G.~V. Eleftheriades, ``Full-duplex nonreciprocal beam steering
  by time-modulated phase-gradient metasurfaces,'' \emph{{Phys. Rev. Appl.}},
  vol.~14, no.~1, p. 014027, 2020.

\bibitem{salary2020time}
M.~M. Salary and H.~Mosallaei, ``Time-modulated conducting oxide metasurfaces
  for adaptive multiple access optical communication,'' \emph{{IEEE Trans.
  Antennas Propagat.}}, vol.~68, no.~3, pp. 1628--1642, 2020.

\bibitem{taravati2020four}
S.~Taravati and G.~V. Eleftheriades, ``Four-dimensional wave transformations by
  space-time metasurfaces,'' \emph{arXiv preprint arXiv:2011.08423}, 2020.

\bibitem{wang2020nonreciprocity}
X.~Wang, G.~Ptitcyn, A.~D{\'\i}az-Rubio, V.~S. Asadchy, M.~S. Mirmoosa, S.~Fan,
  and S.~A. Tretyakov, ``Nonreciprocity in bianisotropic systems with uniform
  time modulation,'' \emph{arXiv}, pp. arXiv--2001, 2020.

\bibitem{nagulu2020multi}
A.~Nagulu, T.~Chen, G.~Zussman, and H.~Krishnaswamy, ``Multi-watt, 1-{GH}z
  {CMOS} circulator based on switched-capacitor clock boosting,'' \emph{IEEE J.
  Solid-State Circuits}, vol.~55, no.~12, pp. 3308--3321, 2020.

\bibitem{wu2020frequency}
X.~Wu, M.~Nafe, A.~{\'A}. Melc{\'o}n, J.~S. G{\'o}mez-D{\'\i}az, and X.~Liu,
  ``Frequency tunable non-reciprocal bandpass filter using time-modulated
  microstrip $\lambda$g/2 resonators,'' \emph{IEEE Transactions on Circuits and
  Systems II: Express Briefs}, 2020.

\bibitem{zang2019nonreciprocal}
J.~Zang, A.~Alvarez-Melcon, and J.~Gomez-Diaz, ``Nonreciprocal phased-array
  antennas,'' \emph{{Phys. Rev. Appl.}}, vol.~12, no.~5, p. 054008, 2019.

\bibitem{Taravati_LWA_2017}
S.~Taravati and C.~Caloz, ``Mixer-duplexer-antenna leaky-wave system based on
  periodic space-time modulation,'' \emph{{IEEE Trans. Antennas Propagat.}},
  vol.~65, no.~2, pp. 442 -- 452, Feb. 2017.

\bibitem{Taravati_AMA_PRApp_2020}
S.~Taravati and G.~V. Eleftheriades, ``Space-time medium functions as a perfect
  antenna-mixer-amplifier transceiver,'' \emph{{Phys. Rev. Appl.}}, vol.~14,
  no.~5, p. 054017, 2020.

\bibitem{bhandare2005novel}
S.~Bhandare, S.~K. Ibrahim, D.~Sandel, H.~Zhang, F.~Wust, and R.~No{\'e},
  ``Novel nonmagnetic 30-db traveling-wave single-sideband optical isolator
  integrated in {III/V} material,'' \emph{IEEE J. Sel. Top. Quantum Electron.},
  vol.~11, no.~2, pp. 417--421, 2005.

\bibitem{fang2013experimental}
K.~Fang, Z.~Yu, and S.~Fan, ``Experimental demonstration of a photonic
  aharonov-bohm effect at radio frequencies,'' \emph{{Phys. Rev. B}}, vol.~87,
  no.~6, p. 060301, 2013.

\bibitem{Wang_TMTT_2015}
J.-F. Chang, J.-C. Kao, Y.-H. Lin, and H.~Wang, ``Design and analysis of
  24-{GH}z active isolator and quasi-circulator,'' \emph{{IEEE Trans. Microw.
  Theory Techn.}}, vol.~63, no.~8, pp. 2638--2649, 2015.

\bibitem{correas2019plasmonic}
D.~Correas-Serrano, N.~K. Paul, and J.~S. Gomez-Diaz, ``Plasmonic and photonic
  isolators based on the spatiotemporal modulation of graphene,'' in
  \emph{Micro-and Nanotechnology Sensors, Systems, and Applications XI}, vol.
  10982.\hskip 1em plus 0.5em minus 0.4em\relax International Society for
  Optics and Photonics, 2019, p. 109821I.

\bibitem{Taravati_PRAp_2018}
S.~Taravati, ``Giant linear nonreciprocity, zero reflection, and zero band gap
  in equilibrated space-time-varying media,'' \emph{Phys. Rev. Appl.}, vol.~9,
  no.~6, p. 064012, Jun. 2018.

\bibitem{li2019nonreciprocal}
J.~Li, C.~Shen, X.~Zhu, Y.~Xie, and S.~A. Cummer, ``Nonreciprocal sound
  propagation in space-time modulated media,'' \emph{{Phys. Rev. B}}, vol.~99,
  no.~14, p. 144311, 2019.

\bibitem{Halevi_PRA_2009}
P.~H. Jorge R Zurita-S\'{a}nchez and J.~C. Cervantes-Gonz\'{a}lez, ``Reflection
  and transmission of a wave incident on a slab with a time-periodic dielectric
  function $\epsilon(t)$,'' \emph{{Phys. Rev. A}}, vol.~79, p. 053821, May
  2009.

\bibitem{lira2012electrically}
H.~Lira, Z.~Yu, S.~Fan, and M.~Lipson, ``Electrically driven nonreciprocity
  induced by interband photonic transition on a silicon chip,'' \emph{{Phys.
  Rev. Lett.}}, vol. 109, no.~3, p. 033901, 2012.

\bibitem{van1970noise}
A.~Van Der~Ziel, ``Noise in solid-state devices and lasers,'' \emph{{Proc.
  IEEE}}, vol.~58, no.~8, pp. 1178--1206, 1970.

\bibitem{hyde1964varactor}
F.~Hyde, ``Varactor-diode parametric amplifiers,'' in \emph{Proceedings of the
  Institution of Electrical Engineers}, vol. 111, no.~6.\hskip 1em plus 0.5em
  minus 0.4em\relax IET, 1964, pp. 1080--1087.

\bibitem{kita1963low}
S.~Kita and K.~Tahara, ``Low-noise 11-{G}c parametric amplifier using
  refrigerated silver-bonded germanium diode,'' \emph{Proceedings of the IEEE},
  vol.~51, no.~4, pp. 618--619, 1963.

\bibitem{lee2010low}
W.~Lee and E.~Afshari, ``Low-noise parametric resonant amplifier,'' \emph{IEEE
  Transactions on Circuits and Systems I: Regular Papers}, vol.~58, no.~3, pp.
  479--492, 2010.

\bibitem{li2018nonreciprocal}
Y.~S. Li, X.~P. Yu, and Z.~H. Lu, ``Nonreciprocal time-varying transmission
  line with carrier boosting technique for low-noise rf front ends,''
  \emph{{IEEE Microw. Wireless Compon. Lett.}}, vol.~28, no.~11, pp.
  1011--1013, 2018.

\bibitem{Hewlett_Packard}
Hewlett-Packard, ``mm-wave waveguide isolator,'' \emph{Palo Alto, CA, USA,
  R365A Datasheet}, 1986.

\bibitem{Millitech}
Millitech, ``Fullband isolator,'' \emph{Northampton, MA, USA, FBI-28
  Datasheet}, 2009.

\end{thebibliography}

\end{document}